\documentstyle[preprint,aps]{revtex}
\newcommand{\simle}
{\raisebox{-0.75ex}[-1.5ex]{$\;\stackrel{<}{\sim}\;$}}
\newcommand{\simge}
{\raisebox{-0.75ex}[-1.5ex]{$\;\stackrel{>}{\sim}\;$}}
\begin{document}
\draft
\baselineskip 18pt
\title{ Magnetic Susceptibility of the Orbitally Degenerate
( $J= 5/2$ ) Periodic Anderson Model \\
{} - Analysis on the Basis of the Fermi Liquid Theory - }
\author{Hiroshi Kontani}
\address{ Institute for Solid State Physics, University of Tokyo,
 Roppongi 7-22-1, Minato-ku, Tokyo 106}
\author{Kosaku Yamada}
\address{ Department of Physics, Faculty of Science, Kyoto University,
 Kyoto 606-01}
%
\maketitle
\vskip 1cm
\centerline{(Received ~~~~~~~~~~~~~~~~~~~~~~~~~~~~~~~~~~~~~~~)}
\begin{abstract}
\baselineskip 18pt
In the orbitally degenerate ( $J= 5/2$ ) Periodic Anderson Model, 
the magnetic susceptibility is composed of both
the Pauli term and the Van Vleck term, as is well known.
The former is strongly enhanced by the strong correlation between 
$f$-electrons.
But, for the latter, the influence of the strong correlation
has been obscure for years.
In this paper we give the solution of the longstanding problem.
With the aid of the $d=\infty$ approximation, 
we study this problem on the basis of the Fermi liquid theory
with degenerate orbitals, taking account of all the vertex corrections
in a consistent way.
As a result, 
we obtain the simple expression for the magnetic susceptibility,
and show unambiguously that
{\it the Van Vleck term is also highly enhanced} in the strong correlation regime.
This fact explains naturally the enhanced magnetic susceptibility observed in 
many insulating systems ( i.e., Kondo insulator ).
Moreover, we show that the Wilson ratio takes a value around 1 in the metallic system,
in good agreement with experiments.
\end{abstract}
%
%
\vskip 1cm 
\noindent
KEYWORDS \ : \ 
 magnetic susceptibility, Van Vleck susceptibility, 

\noindent
\hskip 0.5cm
 orbitally degenerate ( $J=5/2$ ) Periodic Anderson Model,
 Heavy Fermion, 

\noindent
\hskip 0.5cm
Kondo insulator, Fermi liquid theory, $d=\infty$ approximation

\vskip 2cm

\baselineskip 19pt

\def\b{{\beta}}
\def\s{{\sigma}}
\def\Si{{\Sigma}}
\def\e{{\epsilon}}
\def\k{{ {\bf k} }}
\def\p{{ {\bf p} }}
\def\q{{ {\bf q} }}
\def\x{{ {\bf x} }}
\def\w{{\omega}}
\def\d{{\partial}}
\def\e{{\epsilon}}
\def\l{{\lambda}}
\def\L{{\Lambda}}       
\def\a{{\alpha}}
\def\g{{\gamma}}
\def\G{{\Gamma}}        
\def\v{{\varphi}}
\def\D{{\Delta}}
\def\ddH{{ \frac{\d}{\d H} }}
\def\ddHM{{ \frac1{gM} \frac{\d}{\d H} }}
\def\ddw{{ \frac{\d}{\d \w} }}
\def\dde{{ \frac{\d}{\d \e} }}
\def\i{{ {\rm i} }}
\def\expo{{ {\rm e} }}
\def\dwpi{{ \frac{d \w}{2 \pi \i} }}
\def\depi{{ \frac{d \e}{2 \pi \i} }}
\def\deepi{{ \frac{d \e'}{2 \pi \i} }}

\eject
\section{Introduction}
In this paper, we investigate the uniform magnetic susceptibility, $\chi$,
for the orbitally degenerate ( $J= 5/2$ ) Periodic Anderson Model ( $J= 5/2$ PAM ).
\cite{Zou,Hanzawa_p}
This model is much closer to real heavy Fermion systems
because it includes both the $f$-electron orbital degeneracy
and the $l$-$s$ coupling.
Based on this model, we can explain several characteristic behaviors
observed experimentally, which are unable to be derived from the SU(N)-PAM.
( SU(N)-PAM is familiar for theorists but less realistic model than $J= 5/2$ PAM
in that both the $f$-electron bands and the conduction 
electron bands possess the same N-fold degeneracy. )

One of the characteristic behaviors of the model is the presence of 
the anomalous Hall effect (AHE).
Reference \cite{paper1} clearly explains that 
the anomalous Hall coefficients of heavy Fermion systems are proportional
to the square of the resistivity at low temperatures,
and points out that its mechanism is similar 
to that of AHE on the ferromagnetic metals
proposed by Karplus and Luttinger.
\cite{KL,Sm}

The other characteristic property is the presence of 
the Van Vleck magnetic susceptibility.
\cite{Zhang,Hanzawa_v}
As is well known, the magnetic susceptibility of this model contains the
Van Vleck susceptibility $\chi_V$ 
in addition to the usual Pauli susceptibility $\chi_P$,
so the total magnetic susceptibility is given by $\chi=\chi_P+\chi_V$.
$\chi_P$ comes from the intra-band contribution,
which vanishes in the absence of the Fermi surface,
and $\chi_V$ comes from the inter-band contribution.
Of course, $\chi$ is one of the most important and characteristic 
physical quantities in the heavy Fermion systems.
But, unfortunately, the property of the Van Vleck susceptibility $\chi_V$
remains quite obscure in the presence of the large Coulomb interaction.

Furthermore, its general properties should be studied for understanding
the origin of the strongly enhanced magnetic
susceptibility for {\it so-called} Kondo insulators,
\cite{insulator}
where $\chi_P=0$ and only the Van Vleck susceptibility remains.
We also have to understand the property of $\chi$ in order to
interpret the Knight shift data for super-conductive heavy Fermion systems.
\cite{Ohmi}

Historically, this problem was first studied in ref.
\cite{Zou}.
The authors paid attention only to $\chi_P$, and claimed that
$\chi_V$ could be neglected.
On the contorary, 
mean-field like treatments give the enhanced $\chi_V$ .
\cite{Zhang,Hanzawa_v,meanfield}
But, such approximations
seem to be insufficient to settle this problem because they assume the 
frequency independent renormalization factor, which is a less-realistic assumption.
The settlement of this problem will bring us the useful
understanding for the electronic properties for the
orbitally degenerate model.

The existence of $\chi_V$ in $J=5/2$ PAM is closely related
with the fact that the magnetization operator ${\hat M}$ is {\it not}
conserved in this model, i.e., $[ {\hat H},{\hat M}]\neq0$.
 \cite{Nakano}
Whenever ${\hat M}$ is conserved, $\chi_V$ vanishes identically and
the susceptibility has been obtained without ambiguity
in the framework of the Fermi liquid theory ( FLT ).
For instance, the susceptibility for SU(2)-PAM or $J=5/2$ single-site Anderson
model is made of $\chi_P$ alone, and obtained by the Luttinger's method.
 \cite{Luttinger,Shiba,Yamada_p}
On the other hand, in order to analyze the magnetic susceptibility
of our model, we have to find another useful method.
( By the RPA-approximation, which is an insufficient analysis
for our problem, the enhancement for $\chi_V$ is less than factor of two
even in the strong correlation regime where $\chi_P \simle \infty$.
\cite{RPA} )

In this paper, 
we analyze the properties of $\chi$ and $\chi_V$ by investigating
all the vertex correction for them
by use of a kind of the Bethe-Salpeter equation.
Then we employ the $d=\infty$ approximation,
\cite{d}
which is considerably effective for our problem on $J=5/2$ PAM.
\cite{paper3}
We also discuss the Wilson ratio $R$ in the strong correlation regime.
The definition of $R$ is
\begin{eqnarray}
R \equiv \frac{\chi/\chi^0}{\g/\g^0} \cdot R^0 ,
 \label{eqn:Wilson} 
\end{eqnarray}
where $R^0$ is the unrenormalized Wilson ratio.
$\g$ ( $\g^0$ ) is the ( unrenormalized ) 
$T$-linear coefficient of the specific heat,
 $\chi$ ( $\chi^0$ ) is the ( unrenormalized )
uniform magnetic susceptibility.
$R^0$ is given by
\begin{eqnarray}
R^0 = \frac{3 \chi^0}{2 g^2 \mu_B^2 J_{\rm eff}^2 \rho^0(0)},
 \label{eqn:Wilson2}
\end{eqnarray}
where $\rho^0(0)$, $g$ and $\mu_B$ are the density of states for $f$-electrons,
the Lande's $g$-factor, and the Bohr magneton respectively,
and $J_{\rm eff}^2= J(J+1)$ in the spherical system,
i.e., without any electronic crystal field (ECF).
For instance, for an isolated Kondo atom $R$ is universally given by
$R=(2J+1)/2J$ and $R^0= 1$.

The composition of this paper is as follows :
in \S 2, we explain main properties of $J= 5/2$ PAM and its Green's functions.
In \S 3, we express the magnetic susceptibility
in terms of the dynamical susceptibility on the basis of the 
multicomponent FLT.
In the insulating systems, the susceptibility is given by only
the Van Vleck susceptibility defined in this section.
In \S 4, we calculate $\chi$ and $\chi_V$ explicitly in case $U=0$.
The obtained results are the same as those obtained previously.
In \S 5, we investigate $\chi$ and $\chi_{V(P)}$ in $d=\infty$
in the strong correlation regimes,
by taking all the vertex corrections into account.
Then, in the case of no ECF, 
we find that both $\chi$ and $\chi_V$ are 
proportional to the same enhancement factor.
This implies that
{\it the Van Vleck susceptibility $\chi_V$ is enhanced},
which is the main result of this paper.
In \S 6, we also discuss the Wilson ratio in our system
and find out that $R \sim 1$.
We also point out that the results for $d=\infty$ systems 
are qualitatively correct for the three dimensional system.
In \S 7, the obtained results are summarised 
and the future problems are pointed out.

In Appendix A, we study the magnetic susceptibility for SU(N)-PAM,
and show that the Van Vleck susceptibility is small in this model,
which arises from the difference between 
the magnetic momentum of $f$-electrons and that of conduction electrons.
In Appendix D, we confirm our conclusions numerically,
by use of the ( self-consistent ) second order perturbation treatment 
( ( SC- ) SOPT ).
In Appendix E, we show in detail that $\chi_V''$, 
given by (\ref{eqn:vanvleck2}), vanishes identically.

\section{Model and Green's Function}
In this paper, we study $J= 5/2$ PAM,
which is a realistic model for the Ce-compound heavy Fermion systems
in that it represents the $f$-orbital degeneracy.
In the absence of the magnetic field $H$,
our Hamiltonian is given by
\cite{Hanzawa_p,paper1}
\begin{eqnarray}
& &H= H_0+H_1 \\
& &\ \ H_0 = \sum_{\k \sigma} \epsilon_{\k} c_{\k \sigma}^\dagger c_{\k \sigma}
       + \sum_{\k M} E^f f_{\k M}^\dagger f_{\k M}
       + \sum_{M\k \sigma} ( V_{\k M\sigma}^\ast f_{\k M}^\dagger c_{\k \sigma}
         + V_{\k M\sigma} c_{\k \sigma}^\dagger f_{\k M} )  ,
 \label{eqn:hint1}  \nonumber \\
& &\ \ H_1 = \frac U2 \sum_{\k \k ' \q M\neq M'}
 f_{\k - \q M}^\dagger f_{\k '+ \q M'}^\dagger f_{\k 'M'} f_{\k M} ,
 \label{eqn:hint2} \nonumber
\end{eqnarray}
where $c_{\k \sigma}^\dagger$ is the creation operator 
of a conduction electron with momentum $\k$ and spin $\s$, and
$f_{\k M}^\dagger$ is the creation operator of an $f$-electron with momentum $\k $
and angular momentum $M$, which is the eigenvalue of $J_z= l_z+s_z$.
As $J=7/2$ states are ignored in our model as is usually done,
the $f$-orbital has six-fold degeneracy,
i.e., $M= \{ 5/2, 3/2, \cdots, -5/2 \}$.
$V_{\k M\s}$ in (\ref{eqn:hint1}) is the
mixing potential between the f-electrons and the conduction electrons,
which is given by
\begin{eqnarray}
& &V_{\k M\s}=(4\pi)^{1/2} \sum_m a_{m\s}^M Y_{l=3}^m
 (\theta_k ,\varphi_k )\cdot V  ,
 \label{eqn:hint3} \\
& &\ \ a_{m\s}^M=-\s \{ (7/2-M \s)/7 \}^{1/2} \delta_{m,M-\s/2}
 \ \ \ \ \ \mbox{for} \ J=5/2  , \nonumber
\end{eqnarray}
where $a_{m\s}^M$ is the Clebsh-Goldan ( C-G ) coefficient and
$Y_{l=3}^m(\theta_k ,\varphi_k )$ is the spherical harmonic function.
In the Coulomb interaction term (\ref{eqn:hint2}),
we take account of the Pauli's principle, i.e., $M\neq M'$.

In the presence of the magnetic field $H$ along $z$-axis,
both $E^f$ and $\e_{\k}$ in (\ref{eqn:hint1}) are
shifted by the Zeeman energies as
\begin{eqnarray}
\left\{
\begin{array}{l}
E_{M}^f =E^f + g \mu_{\rm B} M \cdot H,    \\
\e_{\k\s}= \e_{\k} + \mu_{\rm B} \s \cdot H,
\end{array}
\right.
\end{eqnarray}
where $\mu_{\rm B}$ is the Bohr magneton
and $\s=1$ for up-spin and $\s=-1$ for down-spin.
$g$ is the Lande's $g$-factor ( $g=6/7$ for $J=5/2$ ).
Hereafter, we put $\mu_B=1$.

The general forms of the Green's functions for $J=5/2$ PAM 
in the absence of the magnetic field are given by 
(3.11) $\sim$ (3.13) of ref. \cite{paper1}.
In the presence of the magnetic field $H$,
the $f$-electron Green's functions can be expressed as
\begin{eqnarray}
& & G_{\k MM'}(\w) 
= G_{\k MM'}^f(\w) + \sum_{\s\s'} \a_{\k M\s}^\ast(\w) 
 G_{\k \s\s'}(\w) \a_{\k M'\s'}(\w), \\
& &G_{\k MM'}^f(\w) \equiv { \left(\w \hat 1 +\mu \hat 1 - {\hat E}^f - 
 {\hat \Sigma}_\k (\w) \right) }_{MM'}^{-1}, \\
& &\a_{\k M\s}\equiv \sum_{M'} V_{\k M'\s}G_{\k M'M}^f(\w), 
 \label{eqn:alpha}
\end{eqnarray}
where $\mu$ denotes the Fermi energy of this system
and $G_{\k \s\s'}(\w)$ is the conduction electron Green's function.
Considering the Zeeman terms for only the $f$-electrons, 
we can derive the expression for the conduction electron Green's function
in the presence of the magnetic field :
\begin{eqnarray}
G_{\k \s\s'}(\w)= 
 \frac{\delta_{\s \s'}\cdot (\w+\mu-\e_{\k})+ (1-2\delta_{\s \s'})
   S_{\k \s\s'}(\w) }
 {(\w+\mu-\e_{\k}-S_{\k\uparrow \uparrow}(\w))
  (\w+\mu-\e_{\k}-S_{\k\downarrow \downarrow}(\w))
 -S_{\k\uparrow \downarrow}(\w) S_{\k\downarrow \uparrow}(\w) },
 \label{eqn:Gr}
\end{eqnarray}
where 
\begin{eqnarray}
S_{\k\s\s'}(\w) \equiv \sum_{MM'}V_{\k M\s}G_{\k MM'}^f(\w)V_{\k M'\s'}^\ast.
\end{eqnarray}
%
Note that $S_{\k\s,-\s}(\w) =0$ in case $H=0$.
The pole of $G_{ \k \s}(\w)$ represents the spectrum for the quasiparticle
( i.e., heavy electron ), which we denote as $E_{\k}^\ast$.
The pole of $G^{f}(\w)$ represents the local $f$-electron spectrum,
$E^f$, and never represent the quasiparticle spectrum.
$G_{\k M}(\w)$ contains both poles.

In Fig. 1,
the electronic structure of $J=5/2$ PAM in case $U=0$  
is sketched in terms of the one-body picture.
There remains four-fold degenerate 
local $f$-electron spectrum, which reflects the difference of the
degeneracy between the conduction electrons and the $f$ electrons.
( They do not exist in SU(N)-PAM. )
In Fig. 1(a), the Fermi energy $\mu$ lies below the lower edge of the
$f$-$c$ hybridization gap, so the system is metallic.
$k_F$ is the Fermi momentum and
$D$ ( $-D$ ) is the upper ( lower ) edge of the conduction band.
We can show that $V^2/D^2$ is proportional to $m_{f{\mbox -}{\rm band}}/m_c$,
where $m_{f{\mbox -}{\rm band}}$ and $m_c$ represent the
unperturbed mass of f-electrons and the conduction electrons, respectively.
In usual heavy fermion systems, $V^2/D^2 \ll 1$ is satisfied.
\cite{paper3,Okada}
On the other hand, in Fig. 1(b), the Fermi energy $\mu$ lies 
in the hybridization gap and below $E^f$, so the system is insulating.
$\D_{-}$ ( $\D_{+}$ ) represents the lower ( higher )
edge of the hybridization gap measured from the Fermi energy.
This situation is a prototype for the {\it so-called} Kondo insulator.
\cite{insulator}
More detailed electronic structure is studied beyond the one-body picture
by use of the numerical perturbation calculation.
\cite{paper3}

In the absence of the magnetic field, $G_{k\s\s'}(\w)$ is diagonal
with respect to the spin $\s$ even if $U\neq 0$.
\cite{paper1}
The density of states ( DOS ) of the $f$-electrons, $\rho^0(\w)$, and 
that of the conduction electrons, ${\rho_c}^0(\w)$,
are given by the retarded Green's functions as
\begin{eqnarray}
& &{\rho_c}^0(\w) \equiv - \frac1{\pi N}\sum_\k {\rm Im} {G_{\k\s}^{0R}}(\w),
 \label{eqn:rho_c} \\
& &\rho^0(\w) \equiv - \frac1{\pi N}\sum_\k {\rm Im} {G_{\k M}^{0R}}(\w)
 = \frac{3V^2}{( \w+\mu-E^f )^2} \cdot {\rho_c}^0(\w),
 \label{eqn:rho}
\end{eqnarray}
where $N$ represents the total number of the $f$-sites.
In case $U\neq 0$,  
these relations (\ref{eqn:rho_c}) and (\ref{eqn:rho})
are also satisfied with $E^f$ replaced by $E^f+\Si(\w)$ for $|\w| \simle T^\ast$,
where $T^\ast ( >0 )$ represents the characteristic energy
within which the quasiparticles are well defined.
In usual heavy Fermion systems, $T^\ast$ is much smaller than $D$.

As is shown in ref. \cite{paper1},
$G_{\k MM'}(\w) \propto \expo^{\i (M'-M)\v_k}$ even in case $U\neq0$,
which is an important relation throughout this paper.

\section{Definition of the Pauli and Van-Vleck Susceptibility}
In this section, we consider both the Pauli susceptibility $\chi_P$ and the
Van Vleck susceptibility $\chi_V$, and derive their general expressions
on the basis of the FLT for $J=5/2$ PAM.
Though our discussion in this section is restricted to $J=5/2$ PAM
( in the presence of any ECF ) for simplicity,
it is more general and is valid for many kinds of
multicomponent Fermi liquid systems.
Now, we consider the situation where the system is in a finite magnetic field
$H$ along $z$-axis.
The magnetization of this system, $\langle M \rangle$,
is expressed by use of the Green's functions as
\begin{eqnarray}
\langle M \rangle
 = \frac1N \sum_\k \int_{\infty}^{\infty} \dwpi \expo^{+\i \w \cdot 0}
  \left\{ \sum_M gMG_{\k MM}(\w)+ \sum_\s \s G_{\k\s\s}(\w) \right\}.
 \label{eqn:M}
\end{eqnarray}
The second term of the above equation is 
the contribution from the spin of the conduction electrons.
In this paper we neglect this contribution because it is very small.
The total susceptibility $\chi$ is given by
\begin{eqnarray}
\chi= \left. \frac{\d \langle M \rangle}{\d H} \right|_{H=0}.
 \label{eqn:chi_ddh}
\end{eqnarray}

As is well known, the Green's function $G_{k\s\s}(\w)$ is expressed by the retarded
and the advanced Green's functions, $G_{k\s\s}^R(\w)$ and $G_{k\s\s}^A(\w)$ as
\begin{eqnarray}
G_{\k\s\s}(\w)= G_{\k\s\s}^R(\w) \cdot \theta(E_{\k\s}^\ast-\mu)
 + G_{\k\s\s}^A(\w) \cdot \theta(\mu-E_{\k\s}^\ast)
 \label{eqn:gAgR}
\end{eqnarray}
for $|k| \sim k_{\rm F}$ and $\w \sim 0$.
\cite{AGD}
$E_{k\s}^\ast$ is the quasiparticle spectrum obtained by
\begin{eqnarray}
\{ G_{\k\s\s}(E_{\k\s}^\ast) \}^{-1}=0.
\end{eqnarray}
From the definition of $E_{k\s}^\ast$ above, we can show 
for $|\k|=k_{\rm F}$ that
\begin{eqnarray}
\ddH E_{\k\s}^\ast 
 = a_\k(0) \cdot \sum_{MM'} \a_{\k M\s}(0)
 \left( M \delta_{M, M'} + \ddH \Si_{\k M M'}(0) \right)
 \a_{\k M'\s}^\ast(0),
 \label{eqn:ddHE}
\end{eqnarray}
where $\a_{\k M\s}(\w)$ is introduced by (\ref{eqn:alpha}), 
and $a_\k(\w)$ is the renormalization factor for the conduction
electrons, given by

\begin{eqnarray}
1/a_\k(\w)= 
 \sum_{MM'} \a_{\k M\s}(\w) \cdot \left( \delta_{MM'}-\ddw \Si_{\k MM'}(\w) \right)
 \cdot \a_{\k M'\s}^\ast(\w).
\end{eqnarray}
Note that $a_\k(\w) \ll 1$ for $|\w| \simle T^\ast$. 
By use of (\ref{eqn:gAgR}) and (\ref{eqn:ddHE}), we can derive that
\begin{eqnarray}
& &-\ddw G_{\k MM'}(\w)= \sum_{M''M'''} \v_{\k MM',M''M'''}(\w) \cdot 
 \left( \delta_{M''M'''}-\ddw \Si_{\k M'''M''}(\w) \right), \\
& &\ddH G_{\k MM'}(\w)= \sum_{M''M'''} \left\{ \v_{\k MM',M''M'''}(\w)
 + Q_{\k MM',M''M'''}(\w) \right\} \nonumber \\
& & \ \ \ \ \ \ \ \ \ \ \ \ \ \ \ \ \ \ \ \ \ \ \ \ \ \ \ \ \ 
 \times \left( M'' \delta_{M''M'''}+\ddH \Si_{\k M'''M''}(\w) \right),
 \label{eqn:ddHG}
\end{eqnarray}
where $\v$ and $Q$ are defined by
\begin{eqnarray}
& &\v_{\k MM',M''M'''}^{R(A)}(\w)= 
 \{ G_{\k MM'''}(\w) G_{\k M''M'}(\w) \}^{R(A)}, \\
& &Q_{\k MM',M''M'''}(\w)= -2 \pi \i a_\k(0)^2 \delta(\w)\delta(E_\k^\ast)
 \cdot \sum_{\s\s'}\a_{\k M\s}^\ast(0)\a_{\k M'\s'}(0)
 \a_{\k M''\s'}^\ast(0)\a_{\k M'''\s}(0), 
 \label{eqn:Q} 
\end{eqnarray}
where the superscript $R(A)$ represents retarded ( advanced ) function.
On the other hand, we can show that
\begin{eqnarray}
& &\lim_{\k\rightarrow \k'} \lim_{\w\rightarrow \w'}
 G_{\k MM'''}(\w) G_{\k' M''M'}(\w') = Q_{\k MM',M''M'''}(\w)+ \v_{\k MM',M''M'''}(\w), 
 \label{eqn:k_lim} \\
& &\lim_{\w\rightarrow \w'} \lim_{\k\rightarrow \k'}  
 G_{\k MM'''}(\w) G_{\k' M''M'}(\w') = \v_{\k MM',M''M'''}(\w),
\end{eqnarray}
on the basis of the well-known FLT.
\cite{Nakano,AGD,Nozier}
By use of (\ref{eqn:M}), (\ref{eqn:chi_ddh}), (\ref{eqn:ddHG}) 
and (\ref{eqn:k_lim}), we can prove that
\begin{eqnarray}
\chi = \lim_{|\k|\rightarrow 0} \lim_{\w\rightarrow0}\chi_\k(\w).
 \label{eqn:k_limit}
\end{eqnarray}
This relation is trivial whenever the magnetization operator is {\it conserved}
\cite{Kubo,Nozier}
but should be proved for our model.
 \cite{Nakano}
Anyway, by use of (\ref{eqn:k_limit}),
 we can discuss our problem in terms of the more familiar FLT.
\cite{Nakano,AGD,Nozier}
So, $\chi$ can be expressed more explicitly as
\begin{eqnarray}
& &\chi= \frac 1N \sum_\p \int \depi {\rm Tr} \left\{
 {\hat M} \left( {\hat \v}_\p(\e) + {\hat Q}_\p(\e) \right)
 {\hat \L}_\p^k(\e) \right\}, 
 \label{eqn:total} \\
& &\ \ \ {\hat \L}_\p^k(\e) \equiv
 {\hat M}+ \frac 1N \sum_\q \int \deepi \left\{ {\hat \G}_{\p\q}^k (\e,\e')
 \left( {\hat \v}_\q(\e') + {\hat Q}_\q(\e') \right) {\hat M} \right\}
 \nonumber \\
& &\ \ \ \ \ \ \ \ \ \ \ \ \ \ 
 \left( \ \  = {\hat M}+ \ddH {\hat \Si}_\p(\e) \ \ \right) , \nonumber
\end{eqnarray}
where Tr represents the trace with respect to the angular momentum,
and $\{ { \hat M } \}_{M_1 M_2}= M_1 \cdot \delta_{M_1 M_2}$.
Here, we have introduced the $k$-limit and $\w$-limit of 
the four-point vertex ${\hat \G}_{\p\p',\q}(\e,\e';\w)$,
and represents them as ${\hat \G}_{\p\p'}^k(\e,\e')$ and ${\hat \G}_{\p\p'}^\w(\e,\e')$,
respectively :
\begin{eqnarray}
& &{\hat \G}_{\p\p'}^k(\e,\e') \equiv
 \lim_{|\q|\rightarrow0} \lim_{\w\rightarrow0}{\hat \G}_{\p\p',\q}(\e,\e';\w), \\
& &{\hat \G}_{\p\p'}^\w(\e,\e') \equiv
 \lim_{\w\rightarrow0} \lim_{|\q|\rightarrow0}{\hat \G}_{\p\p',\q}(\e,\e';\w),
\end{eqnarray}
where ${\hat \G}_{\p\p'}^{k(\w)}(\e,\e')$ expresses the matrix whose
$(MM',M''M''')$-component is $\G_{\p\p',MM',M''M'''}^{k(\w)}(\e,\e')$.
They are expressed by Fig. 2.

As is well known, ${\hat \G}_{\p\p'}^k(\e,\e')$ and ${\hat \G}_{\p\p'}^\w(\e,\e')$
are related to each other by the following Bethe-Salpeter equation :
\cite{AGD,Nakano} 
\begin{eqnarray}
{\hat \G}_{\p\p'}^k(\e,\e') = {\hat \G}_{\p\p'}^\w(\e,\e') 
 + \frac 1N \sum_\q \int \dwpi  \left\{ {\hat \G}_{\p\q}^k(\e,\w) {\hat Q}_\q(\w)
 {\hat \G}_{\q\p'}^\w(\w,\e') \right\}.
\end{eqnarray}
For instance, we can write the $H$-derivative and the $\w$-derivative 
of the selfenergy as follows :
\begin{eqnarray}
& &-\ddw {\hat \Si}_\p(\e)= \frac 1N \sum_\q \int \dwpi
 \left\{{\hat \G}_{\p\q}^\w(\e,\w){\hat \v}_\q(\w) \right\},\\
& &\ddH {\hat \Si}_\p(\e)= \frac 1N \sum_\q \int \dwpi 
 \left\{{\hat \G}_{\p\q}^k(\e,\w) \left( {\hat Q}_\q(\w)+{\hat \v}_\q(\w)
 \right) {\hat M} \right\}.
\end{eqnarray}

In this paper, we define the Van Vleck susceptibility $\chi_V$
and the Pauli susceptibility $\chi_P$ as
\begin{eqnarray}
& &\chi_V \equiv \lim_{\w\rightarrow0} \lim_{|\k|\rightarrow0} \chi_\k(\w),
 \label{eqn:def_v} \\
& &\chi_P \equiv \chi-\chi_V.
 \label{eqn:def_p}
\end{eqnarray}
According to (\ref{eqn:def_v}) and (\ref{eqn:def_p}),
we can divide $\chi$ into $\chi_P$ and $\chi_V$ in a unique way.
The standard FLT tells us that $\chi_V=0$ and $\chi=\chi_P$
if the magnetization operator ${\hat M}$ is conserved, i.e., 
$[{\hat H},{\hat M}]=0$ :
 \cite{Nozier,Nakano} 
we briefly show this in Appendix A.
Equations (\ref{eqn:def_v}) and (\ref{eqn:def_p}) can be expressed as
\begin{eqnarray}
& &\chi_V= \frac 1N \sum_\p \int \depi {\rm Tr} \left\{
 {\hat M}{\hat \v}_\p(\e) {\hat \L}_\p^\w(\e) \right\}, 
 \label{eqn:vanvleck} \\
& &\chi_P= \frac 1N \sum_\p \int \depi {\rm Tr} \left\{
 {\hat \L}_\p^\w(\e){\hat Q}_\p(\e) {\hat \L}_\p^k(\e) \right\},
 \label{eqn:pauli} \\
& &\ \ \ \ {\hat \L}_\p^\w(\e) \equiv {\hat M}+ 
 \frac 1N \sum_\q \int \deepi \left\{ {\hat \G}_{\p\q}^\w (\e,\e')
 {\hat \v}_\q(\e') {\hat M} \right\} ,
 \label{eqn:Lam_k} \\
& &\ \ \ \ \ \ \ \ 
 \left( \ \ =  {\hat M}+ 
 \left. \ddH {\hat \Si}_\p(\e) \right|_{\w{\mbox -}{\rm limit} }
 \neq {\hat \L}_\p^k(\e). \ \ \right) .  \nonumber
\end{eqnarray}
Equation (\ref{eqn:vanvleck}) is depicted by Fig. 3.
In the following sections, we investigate the expressions of $\chi$ and $\chi_V$, 
given by (\ref{eqn:total}) and (\ref{eqn:vanvleck}), further.

Here we stress that the principle of division of the total susceptibility 
into (\ref{eqn:def_v}) and (\ref{eqn:def_p}) is very clear ;
the former does not contain any ${\hat Q}_p(\w)$ defined by (\ref{eqn:Q})
while the latter contains at least one ${\hat Q}_p(\w)$ in itself.
Equation (\ref{eqn:pauli}) indicates that $\chi_P$ vanishes
and the total susceptibility is given by $\chi_V$
in the case where ${\hat Q}_p(\e) \equiv 0$, i.e., where the Fermi surface
disappears for some reasons or other.
This is true for any multicomponent Fermi liquid systems.


\section{Magnetic Susceptibility $\chi^0$ : in Case $U=0$}
In this section, we calculate both $\chi_V^0$ and $\chi_P^0$ 
defined by (\ref{eqn:vanvleck}) and (\ref{eqn:pauli}) in case $U=0$,
and show that our results are identical to those obtained 
previously by several authors.
\cite{Zhang,Hanzawa_v}

\begin{eqnarray}
& &\chi^0= -\frac1N \sum_{\k MM'}\int\dwpi M \left\{
 \v_{\k,MM,M'M'}^0(\w)+ Q_{\k,MM,M'M'}^0(\w) \right\}M', 
 \label{eqn:chi0}\\
& &\chi_V^0= -\frac1N \sum_{\k MM'}\int\dwpi M \v_{\k,MM,M'M'}^0(\w) M'.
 \label{eqn:chiV0}
\end{eqnarray}

a) Metallic case \ : \
For simplicity, we consider the case where the DOS for the
conduction electrons is constant, i.e., $\rho_c^0(\w)= \rho_c$.
We also neglect the contribution from
the spin of the conduction electrons, which is of order 
$\sim \rho_c/\rho \sim O(V^2/D^2)$ 
compared with the total magnetic susceptibility.
 ( see \S 2. ).
Moreover, we assume the six-fold degenerate $f$-electron 
spectrum and the spherical Fermi surface.
The calculated results of (\ref{eqn:chi0}) and (\ref{eqn:chiV0}) are
\begin{eqnarray}
& &\chi^0= 2g^2J(J+1)\cdot A\rho^0(0),
 \label{eqn:susc0} \\
& &\chi_V^0 = 2g^2J(J+1)\cdot C\rho^0(0), 
 \label{eqn:vanvleck0} \\
& &\ \ A= 0.333, \ \ C= 0.152. \nonumber
\end{eqnarray}
Note that $\rho^0(\w)$, given by (\ref{eqn:rho}), is the DOS for $f$-electrons.
Moreover, the ratio
$\chi_P^0/\chi_V^0 = (A-C)/C = 1.19$
is derived.
The Wilson ratio $R^0$ of this system,
whose definition is given by (\ref{eqn:Wilson}), is given by
\begin{eqnarray}
R^0=1.  \label{eqn:R0}
\end{eqnarray}
The obtained results (\ref{eqn:susc0}) $\sim$ (\ref{eqn:R0})
are identical to those shown previously by several authors.
\cite{Zhang,Hanzawa_v}
Note that these results 
are not universal for $J=5/2$ PAM
in that it depends on the $\w$-dependence of $\rho_c(\w)$.

b) \ Insulating Case \ : \
Next, we consider the insulating case where the Fermi energy $\mu$
lies in the $f$-$c$ hybridization gap.
In this case, apparently $\chi_P^0=0$ because $\rho^0(0)=0$.
So, the total magnetic susceptibility $\chi^0$ is equal to
the Van Vleck susceptibility, which is given by
\begin{eqnarray}
\chi_{\rm ins}^0 &=&
2g^2J(J+1) \cdot C\rho^0(\D_{-}),
 \label{eqn:vanvleck00} \\
\rho^0(\D_{-}) &=& \rho_c \cdot \frac{3V^2}{(\D_{-}+\mu-E^f)^2} \ >0,
 \nonumber
\end{eqnarray}
where $\D_{-}$ is the lower edge of the hybridization gap.
In this case, $\chi_{\rm ins}^0$ remains finite and will take a large value.
This result makes highly contrast to that of the 
insulating SU(6)-PAM, whose magnetic susceptibility is about 
$V^2/D^2$ times smaller than (\ref{eqn:vanvleck00}).
( see Appendix A. )

\section{Magnetic Susceptibility $\chi$ in the $d=\infty$ limit : 
 in Case $U \neq 0$}

In this section, we analyze the magnetic susceptibility $\chi$
in the strong coupling regime, taking account of all the vertex corrections
due to the strong Coulomb interaction $U$ in a consistent way.
In \S 5 and \S 6, we assume the six-fold degeneracy of $f$-electron spectrum
and spherical Fermi surface in the extended zone scheme.
Such a assumption will be allowed for our aim to elucidate
the essential properties of the susceptibility under the influence of
the strong Coulomb interaction.

We study this problem concerning only the leading term with respect to
$1/d$-expansion ;
we call such an approximation {\it the $d=\infty$ approximation}
or {\it the $d=\infty$ limit}.
In the $d=\infty$ limit, the selfenergy becomes momentum-independent,
as is well known.
\cite{d}
For $J=5/2$ PAM, the electronic properties obtained by this approximation will
be quite similar to that in the realistic $d=3$ system.
\cite{paper3}
So, the conclusions obtained by the $d=\infty$ limit analysis 
will be valid also for the $d=3$ system.

\subsection{Brief review of the $d=\infty$ approximation} 

In the $d=\infty$ approximation, 
the irreducible four-point vertex is considered as {\it local},
as well as the selfenergy.
So, the reducible four-point vertex $\G$
is composed of the {\it local} irreducible four-point vertex $\G^I$
and the {\it non-local} particle-hole Green's functions ${\hat \v}(\w)$,
as is shown in Fig. 4.
We explain the reasoning briefly :

Fig. 5 shows the two examples of $U^2$-order contributions for the susceptibility.
We assume the $d$-dimensional hyper-cubic $f$-electron lattice for instance.
Each site has $z=2d$ nearest neighbor sites.
$j$, $l$, $m$ and $n$ represents the $f$-electron sites,
and we have to take summation over them.
Note that an $f$-electron on one site cannot hop to other sites
without mixing with the conduction electrons.
In Fig. 5 (a), $(j,m)$ and $(l,n)$ are connected
by the Green's functions, while they are not in Fig. 5 (b).
Here, we fix $l$ and $m$ as the $\xi$-th nearest neighbor sites.
Such selections exist about $\sim z^\xi \sim d^\xi$ different ways in total.
In the spirit of $1/d$-expansion,
an itinerant Green's function in the real space connecting 
the $\xi$-th nearest neighbor sites is scaled as $\sim O(d^{-\xi/2})$.
After the summation over $j$, $n$, $l$ and $m$, 
while $l$ and $m$ are limited to the $\xi$-th nearest neighbor pairs,
Fig. 5 (a) and (b) becomes of order $\sim d^\xi$ and $\sim d^0$, respectively.
Thus in the $d=\infty$ limit,
the irreducible four-point vertices,
included in Fig 5 (a) for instance, are local and momentum-independent,
as well as the selfenergy.
On the other hand, the particle-hole pair should be regarded as non-local.
Below, we take account of $O(d^0)$ terms at most, i.e.,
{\it the $d=\infty$ approximation}.

In this approximation, $k$-summation becomes 
\begin{eqnarray}
& &\frac 1N \sum_\k \ \rightarrow \ \int {\rho_c}^0(\e_k) d \e_k,
 \label{eqn:summation} 
\end{eqnarray}
where ${\rho_c}^0(\e_k)$ is the unperturbed DOS for the conduction electrons.
We see below that the results of this paper are
sensitive to the value of ${\rho_c}^0(0)$,
but insensitive to the energy-dependence 
of ${\rho_c}^0(\e_k)$ in heavy Fermion systems.
So, we do not pay attention to the functional form of ${\rho_c}^0(\e_k)$.

\vskip 1cm
\noindent
\subsection{The local Green's function, the selfenergy and
 the irreducible four point vertex in the $d=\infty$ case} 

As is shown in the previous subsection, in the $d=\infty$ limit
both the selfenergy and the irreducible four-point vertex are composed
of the local Green's functions, ${\hat g}(\w)$.
When we assume the spherical Fermi surface,
$g(\w)$ is given by
\begin{eqnarray}
& &\frac 1N \sum_\k G_{\k MM'}(\w) \equiv g(\w) \cdot \delta_{MM'}. 
 \label{eqn:lgreen}
\end{eqnarray}
Thus, 
the local Green's function $g(\w)$ is
diagonal with respect to $M$, and it is independent of $M$ in case $H=0$.
( Notice that $G_{\k MM'}(\w)$ is proportional to the
phase factor $\propto \expo^{\i (M'-M) \v_k}$,
\cite{paper1}
which vanishes after the $\k$-summation in case $M \neq M'$. )
This is a remarkable simplification occurring in the $d=\infty$ approximation.
Thus, in the $d=\infty$ limit,
both the selfenergy and the four-point vertex become diagonal 
with respect to $M$ even in a finite magnetic field $H$,
and independent of $M$ in case $H=0$.

For simplicity, we consider the case in the absence of ECF, i.e.,
$E_M^f = E^f$ for all $M$.
In the absence of the magnetic field $H$, 
we can rewrite the $f$- electron Green's function given by (\ref{eqn:Gr}),
into a simpler form as follows :
\begin{eqnarray}
G_{\k MM'} (\w)= h_{\k MM'}G_\k(\w)+ d_{\k MM'}G^{f}(\w),
\end{eqnarray}
where
\begin{eqnarray}
\left\{
\begin{array}{l}
G_\k(\w) = \left({\w+\mu-E^f-\Si(\w)- 3V^2/(\w+\mu-\e_\k)} \right)^{-1},  \\
G^{f}(\w)= \left({\w+\mu-E^f-\Si(\w)} \right)^{-1},  \\ 
\ \ h_{\k MM'} = {\sum_\s V_{\k M\s}^\ast V_{\k M'\s}}/{3V^2}, \\
\ \ d_{\k MM'} = \delta_{MM'}- h_{\k MM'},
 \label{eqn:green2}
\end{array}
\right.
\end{eqnarray}
where $\Si(\w)$ represents the selfenergy in the $d=\infty$ limit,
which depends on neither $k$ nor $M$.

Here, we consider the quasiparticle representation of 
$G_k(\w)$ given by (\ref{eqn:green2}).
For $|\w| \simle T^\ast$, the retarded Green's function is
\begin{eqnarray}
& &G_\k^R(\w) = \frac{ z_1(\w)}{ \w-E_\k^\ast+ \i \delta }, \\
& &\ \ \ \
 \left( \ {\rm i.e.,} \ \  \left( -1/\pi \right)
 {\rm Im}G_\k^R(\w) = z_1(\w) \cdot \delta(\w-E_\k^\ast) 
 \ \right), \nonumber 
\end{eqnarray}
where $T^\ast$ represents the characteristic energy,
within which the quasiparticle can be well defined.
( $T^\ast$ corresponds to the {\it renormalized} Fermi energy. )
Here, 
$z_1(\w)$ is the momentum-independent 
renormalization factor of $f$-electrons, given by
\begin{eqnarray}
& &z_1(\w) \equiv { \left( 1-\ddw \Si(\w)+ \frac{3V^2}{(\w+\mu-\e_k)^2} 
 \right) }^{-1}, \\
& &z(\w) \equiv { \left( 1-\ddw \Si(\w) \right) }^{-1}.
\end{eqnarray}
Both $z(\w)$ and $z_1(\w)$ are the renormalization factors of $f$-electrons.
Apparently, $1/z_1(0) \cong 1/z(0) \gg 1$ in the heavy Fermion system.

Finally, in the $d=\infty$ limit,
we can show that the irreducible parallel four-point vertex
$\G_{MM,M'M'}^I(\w,\w')$ satisfies the following property :
\begin{eqnarray}
\G_{MM,M'M'}^I(\w,\w') = \G^I(\w,\w') \cdot \delta_{MM'},
 \label{eqn:diag}
\end{eqnarray}
i.e., $\G_{MM,M'M'}^I(\w,\w')$ is diagonal with respect to $M$
and independent of $M$ ( see Fig. 6. )
This property is explained in Appendix. B,
which is no more true for reducible vertices.

Here, we summarise the results of this subsection : 
in our $J=5/2$ PAM, in the $d=\infty$ limit,
the angular momentum $M$ is conserved in
the local Green's function, the selfenergy and the irreducible four
point vertex, because they are estimated as local processes in the $d=\infty$ limit.

\vskip 1cm
\noindent
\subsection{Vertex Corrections for the magnetic susceptibility}

Contrary to $\Si(\e)$ and $\G^I(\e,\e')$, the particle-hole
Green's functions are never restricted to the local process,
so ${\hat \v}(\w)$ and ${\hat Q}(\w)$ are not diagonal with
respect to $M$.
In $d=\infty$, ${\hat \v}(\w)$ is given by
\begin{eqnarray}
{\hat \v}(\w)= \frac 1N\sum_j {\hat G}_{i,j}(\w) \cdot {\hat G}_{j,i}(\w)
 = \frac 1N \sum_\k {\hat G}_{\k}(\w) \cdot {\hat G}_{\k}(\w),
 \label{eqn:vgeneral}
\end{eqnarray}
which is a matrix with respect to $M$.
In  (\ref{eqn:vgeneral}),
${\hat G}_{i,j}(\w)$ ( ${\hat G}_{\k}(\w)$ )
is the Green's function in the real- ( momentum- ) space representation.

Here, we introduce a $M$-independent
particle-hole Green's function in $d=\infty$, $\v_a(\w)$,
\begin{eqnarray}
\v_a(\w) &\equiv& \frac1N \sum_\k \sum_{M'} \v_{\k,MM,M'M'}(\w) \cdot 1 
 \nonumber \\
&=&
 (1-A) \frac1N \sum_\k {G_\k^f}^2(\w)+ A\frac1N \sum_\k G_\k^2(\w),
 \label{eqn:ddw_green}
\end{eqnarray}
where
\begin{eqnarray}
A \equiv 1/3 = \int \frac{d \Omega_\k}{4\pi} \cdot {h_{\k MM}}.
\end{eqnarray}
$\v_a(\w)$ is related with frequency-derivative of the selfenergy,
which is depicted by Fig. 7 (a).
Note that the summation on $M'$ is taken in (\ref{eqn:ddw_green}).

Next, we also introduce another $M$-independent
particle-hole Green's functions, $\v_b(\w)$,
\begin{eqnarray}
\v_b(\w) &\equiv& 
 \frac 1M \left\{ \frac 1N \sum_\k \sum_{M'} \v_{\k,MM,M'M'}(\w)
 \cdot M' \right\}  \nonumber \\
&=&
 (1-2A+B)\frac1N \sum_\k {G_\k^f}^2(\w)+ B \frac1N \sum_\k G_\k^2(\w)  \nonumber \\
&+& 2(-A+B)\frac1N \sum_\k \left( \frac{\w+\mu-\e_\k}{3V^2} \right) G_\k^f(\w)
 \ + \ 2(A-B)\frac1N \sum_\k \left( \frac{\w+\mu-\e_\k}{3V^2} \right) G_\k(\w),
 \label{eqn:ddh_green}
\end{eqnarray}
where
\begin{eqnarray}
& &B \equiv 0.181 = \int \frac{d \Omega_\k}{4\pi} \cdot {j_{\k MM}}/M. \\
& &\ j_{\k MM'} = \sum_{\s \s' M''} V_{\k M\s}^\ast V_{\k M''\s} M''
 V_{\k M''\s'}^\ast V_{\k M'\s'}/(3V^2)^2.
\end{eqnarray}
$\v_b(\w)$ is related with the
magnetic field-derivative of the selfenergy, 
which is depicted by Fig. 7 (b).

We also define the following functions,
which are $M$-independent :
\begin{eqnarray}
& &Q_\a(\w) \equiv \frac 1N \sum_\k \sum_{M'} Q_{\k,MM,M'M'}(\w) 
 = -A \cdot 2 \pi \i z_1(0) \rho(0) \delta(\w) \\
& &Q_\b(\w) \equiv \frac 1M \left\{
 \frac 1N \sum_\k  \sum_{M'} Q_{\k,MM,M'M'}(\w)\cdot M' \right\}
 = -B \cdot 2 \pi \i z_1(0) \rho(0) \delta(\w)  \\
& &Q_\g(\w) \equiv Q_\a(\w) - Q_\b(\w)
 = -C\cdot 2\pi \i z_1(0) \rho(0) \delta(\w), \label{eqn:c_a_b} \\
& &\ \  \left( \ {\rm i.e.,} \ \ C=A-B= 0.152 \ \right) , \nonumber
\end{eqnarray}
where we have used the relation 
$d E_\k^\ast/d \e_\k = a_\k(0)$ and
$z_1(0)= a_\k(0) \cdot 3V^2/(\mu-E^f-\Si(0))^2$ for $|\k|=k_{\rm F}$,
which is rigorous because our selfenergy is $k$-independent.
Note that $\rho(0)$ is not an enhanced quantity.
Especially, in the $d=\infty$ approximation,
$\rho(0)= \rho^0(0)$ is satisfied rigorously 
on condition that $E^f$ is modified so as to fix 
the value of $\mu$,
because of the momentum-independence of the selfenergy.

Here, we notice that $\v_a(\w) \neq \v_b(\w)$ and $Q_\a(\w) \neq Q_\b(\w)$
in our model, which is the consequence of the fact that
the magnetization $M$ is not conserved. i.e., 
$[ {\hat H}, {\hat M} ] \neq 0$.
This situation has been prevented us from treating this problem
on the basis of the FLT.

Then, we introduce the reducible four-point vertices
$\G^{r}(\e,\e')$ and $\G^{r,\xi}(\e,\e')$ so as to 
satisfy the following Bethe-Salpeter equations :
\begin{eqnarray}
\G^{r}(\e,\e') &=& \G^I(\e,\e') +
 \int \dwpi \G^I(\e,\w) \v_{r}(\w) \G^{r}(\w,\e'), 
 \label{eqn:BS1} \\
\G^{r,\xi}(\e,\e') &=& \G^I(\e,\e') +
 \int \dwpi \G^I(\e,\w) ( \v_{r}(\w)+ Q_{\xi}(\w) ) \G^{r,\xi}(\w,\e'),
\end{eqnarray}
where $r=\{ a,a',b \}$ and $\xi= \{ \a,\b,\g \}$.
For example, we can represent the mass-enhancement factor
as follows :
\begin{eqnarray}
& &-\ddw \Si_M(\w) = \int \depi \G^a(\w,\e) \left\{
 \v_a(\e)+ \frac 1N \sum_{\k} G_\k^2(\e) \frac{V^2}{(\e+\mu-\e_k)^2} \right\}
 \nonumber \\
& &\ \ \ \ \ \ \ \ \ \ \ \ \
 \cong \int \depi \G^a(\w,\e) \v_a(\e) ,
 \label{eqn:ddw_S} 
\end{eqnarray}
which is shown in Fig. 8.
Here, we notice that 
$\frac 1N \sum_k \v_{k,MM',M''M''} = 0$
for $M\neq M'$ because of the phase factor, $\expo^{\i(M'-M)\v_k}$.
In the same way, the Van Vleck susceptibility $\chi_V$ 
and the total susceptibility $\chi$, given by (\ref{eqn:vanvleck})
and (\ref{eqn:total}) respectively,
are expressed as
\begin{eqnarray}
\chi_V &\cong& -2g^2 J(J+1) \int \depi \v_{b}(\e) 
 \left\{ 1+ \int \deepi \G^{b}(\e,\e') \v_{b}(\e') \right\} ,
 \label{eqn:chi_v_pre} \\
\chi &\cong& -2g^2 J(J+1) \int \depi (\v_{b}(\e)+ Q_{\b}(\e)) 
 \left\{ 1+ \int \deepi \G^{b,\b}(\e,\e') (\v_{b}(\e')+ Q_{\b}(\e')) 
 \right\} .
 \label{eqn:chi_pre} 
\end{eqnarray}
Notice that \ 
$\sum_{M'M''} \int \! \int \G^I \v_{MM,M'M'} \G^I \v_{M'M',M''M''} M''
 = \sum_{M'} \int \! \int \G^I \v_{MM,M'M'} M' \cdot \G^I \v_b
 = M \cdot \int \! \int \G^I \v_b \G^I \v_b 
 \subset M\cdot \int \G^b \v_b$ \ 
and \ $\sum_M M^2 = 2J(J+1)$ for $J=5/2$.
In (\ref{eqn:ddw_S}) ( (\ref{eqn:chi_v_pre}) and (\ref{eqn:chi_pre}) ), 
we have omitted the contribution
from the ( spins of the ) conduction electrons.
We neglect it hereafter because its 
value is at most of order 
$\sim \rho_c(0)/\rho(0) \sim O(V^2/D^2) \ll 1$
in the real heavy Fermion systems,
\cite{explain}
where $\rho(0)$ ( $\rho_c(0)$ )
are the density of states of $f$-electrons ( conduction electrons )
at the Fermi energy, introduced by (\ref{eqn:rho}) ( by (\ref{eqn:rho_c}) ).

In heavy Fermion systems, the behavior of
${\rm Im}\v_{a} (\e)$ or ${\rm Im}\v_{b} (\e)$ 
only for $0 \leq -\e \simle T^\ast$
plays a predominant role for the enhancement factor,
as is discussed in Appendix C.
Below, we show the relation between ${\rm Im}\v_{a} (\e)$ and ${\rm  Im}\v_{b} (\e)$,
which is derived in Appendix C.

\noindent
(i) In the metallic case :
Here, we introduce the new particle-hole Green's functions, $\v_{a'}(\e)$, as
\begin{eqnarray}
& &\v_{a'}(\e) \equiv \v_b(\e)- Q_\g(\e)
 \label{eqn:abQ}.
\end{eqnarray}
Then, the following important relation is obtained :
\begin{eqnarray}
& &{\rm Im}\v_{a'}(\e) \cong {\rm Im}\v_a(\e)  \ \ \ \ \ \ \ \ 
 {\rm for} \ \ \ 0 \leq -\e \simle T^\ast.
 \label{eqn:important}
\end{eqnarray}

\noindent
(ii) In the insulating case ( i.e., $\rho(0)=0$ ) :
Here, we introduce $\v_{a'} (\e)$ as
\begin{eqnarray}
& &\v_{a'} (\e) \equiv \v_b(\e) - Q_{\g}^i(\e), \\
& &\ Q_{\g}^i (\e) \equiv -C \cdot 2 \pi \i z(\D_{-}^\ast)
 \rho(\D_{-}^\ast) \cdot \delta(\e-\D_{-}^\ast),
 \label{eqn:Qcd}
\end{eqnarray}
where $\D_{-}^\ast$ denotes the renormalized lower edge of the
hybridization gap measured from the fermi energy, and
$\rho(\D_{-}^\ast) = \rho_c \cdot (\D_{-}^\ast+\mu-\e_{k_F})^2/3V^2$.
Apparently, the relation
$\rho(\D_{-}^\ast) \cong \rho^0(\D_{-})$
is satisfied very well.
Because the relation $1/z(\D_{-}^\ast) \cong 1/z(0)$ is naturally expected
by the numerical calculation in Appendix D,
the relation
$\D_{-}^\ast \cong z(0) \cdot \D_{-}$
is also expected.

We can show that
\begin{eqnarray}
{\rm Im}\v_{a'}(\e) \cong {\rm Im}\v_a(\e), \ \ \ \ \ \ \ \ 
 {\rm for} \ \ \ 0 \leq -\e -\D_{-}^\ast + \simle T^\ast.
 \label{eqn:important2}
\end{eqnarray}
%

In the next subsection, we investigate the magnetic susceptibility 
$\chi$ and $\chi_V$ by use of the relation 
(\ref{eqn:important}) and (\ref{eqn:important2}), respectively.

\vskip 1cm
\noindent
\subsection{The expression for the magnetic susceptibility 
 in the $d=\infty$ case. } 

In this subsection, we study the 
(\ref{eqn:chi_v_pre}) and (\ref{eqn:chi_pre}) further
and obtain the simple expressions for 
$\chi$, $\chi_V$ and $\chi_{\rm ins}$.
By use of the relations 
$\v_b= \v_{a'}+ Q_\g$ and $\v_b+ Q_\b= \v_{a'}+ Q_{\a}$ 
given in the previous section, 
we can check the following Bethe-Salpeter equations 
in a straightforward way :
\begin{eqnarray}
& &\G^{b}(\e,\e') = \G^{a'}(\e,\e') +
 \int \dwpi \G^{a'}(\e,\w) Q_\g(\w) \G^{b}(\w,\e'), 
 \label{eqn:BSab1} \\
& &\G^{b,\b}(\e,\e') = \G^{a'}(\e,\e') +
 \int \dwpi \G^{a'}(\e,\w) Q_{\a}(\w) \G^{b,\b}(\w,\e').
\end{eqnarray}
The first line is depicted in Fig. 9.
Taking account of the relations (\ref{eqn:important}) 
and (\ref{eqn:c_a_b}) obtained in \S 5-3,
we can estimate as
\begin{eqnarray}
\left\{
\begin{array}{l}
\G^{a}(\e,\e') \cong \G^{a'}(\e,\e') \ \ \ \ \ \ \ \ 
 {\rm for} \ \ 0< \{ -\e,-\e' \}\simle T^\ast,   
  \\
\G^{a,\a}(\e,\e') \cong \G^{b,\b}(\e,\e') \ \ \ \ \ 
 {\rm for} \ \ 0< \{ -\e,-\e' \}\simle T^\ast,   
 \label{eqn:ac} 
\end{array}
\right.
\end{eqnarray}
because the contributions to them from the quasiparticle
excitations are dominant
in heavy Fermion systems.
These results play important roles in the following discussions.

Then, we introduce the new enhancement factor, $1/z'(\e)$, as
\begin{eqnarray}
& &\frac 1{z'(\e)} \equiv 1+ \int \deepi \G^{a'}(\e,\e')\v_{a'}(\e').
 \label{eqn:ddh_c}
\end{eqnarray}
Considering the relation (\ref{eqn:ac}),
\begin{eqnarray}
1/z(\e) \cong 1/z'(\e) \ \ \ \ \
 {\rm for} \ \ 0<-\e<\simle T^\ast
 \label{eqn:propor}
\end{eqnarray}
is satisfied in heavy Fermion systems.
In Appendix D, we show that the relation (\ref{eqn:propor}) is well supported by 
the numerical calculation by SOPT and  SC-SOPT.

Here, we introduce the enhancement factor for the magnetic
susceptibility $1/z_H(\e)$ as
\begin{eqnarray}
\frac 1{z_H(\e)} &\equiv& 1+ \left. \ddHM \Si_M(\e) \right|_{H=0}  \\
&=&  1+ \int \deepi \G^{b,\beta}(\e,\e') 
       \left( \v_{b}(\e')+Q_\beta(\e') \right) \nonumber \\
&=& \int \deepi \left( 2\pi\i \delta(\e-\e')+ \G^{b,\b}(\e,\e')Q_{\a}(\e') 
         \right) \frac 1{z'(\e')},
 \label{eqn:z_H}
\end{eqnarray}
where (\ref{eqn:z_H}) is given for the metallic case.
In the insulating case, $\G^{b,\b}$, $Q_{\a}$  in (\ref{eqn:z_H})
is replaced by $\G^{b}$, $Q_{\g}^i$, respectively.
Apparently, $1/z_H(0) \sim 1/z(0)$ is expected.

Then, we can rewrite the total susceptibility $\chi$,
given by (\ref{eqn:chi_pre}), and obtain a simple expression as
\begin{eqnarray}
\chi &=& -2g^2 J(J+1) \int \int \frac{d\e d\e'}{(2\pi\i)^2} \frac 1{z'(\e)} Q_{\a}(\e)
 \left( 2\pi\i \delta(\e-\e')+ \G^{b,\b}(\e,\e')Q_{\a}(\e') \right) \frac 1{z'(\e')}
 + \chi_V'' 
 \nonumber \\
&=& \ \chi^0 \cdot \frac1{z_H(0)} \cdot \frac{z(0)}{z'(0)} \ + \ \ \chi_V'',
 \label{eqn:chi} 
\end{eqnarray}
where $\chi^0 = 2g^2 J(J+1) \cdot A\rho(0)$,
which is similar to the unperturbed value given by (\ref{eqn:susc0}).
And $\chi_V''$ is given by
\begin{eqnarray}
\chi_V''= -2g^2 J(J+1) \int \depi \v_{a'}(\e) \frac 1{z'(\e)}.
 \label{eqn:vanvleck2}
\end{eqnarray}
We can estimate that $\chi_V'' = 0$ by use of (\ref{eqn:propor}), 
because
\begin{eqnarray}
\chi_V'' \cong -\int_{-\infty}^\infty \deepi \v_a(\e)/z_1(\e)
 = \frac 1N \sum_k \int \depi \cdot \frac{\d}{\d \e} G_k(\e) = 0. 
 \label{eqn:zero}
\end{eqnarray}
More accurate derivation of (\ref{eqn:zero}) is given in Appendix E.
Equation (\ref{eqn:chi}) ( and (\ref{eqn:chi_pre}) ) is depicted in Fig. 10.
Thus, by use of (\ref{eqn:propor}) and (\ref{eqn:zero}), 
we get the simple expression,
\begin{eqnarray}
\chi \cong \chi^0 \cdot \frac 1{z_H(0)}.
 \label{eqn:fin-res}
\end{eqnarray}
This is the main result of this paper, and is proved to be {\it rigorous}
when $1/z(\w) \equiv 1/z'(\w)$ is satisfied identically.
( see Appendix D. )

Next, we also rewrite the Van Vleck susceptibility $\chi_V$,
given by (\ref{eqn:chi_v_pre}), as follows,
\begin{eqnarray}
\chi_V = -2g^2 J(J+1) \int \int \frac{d\e d\e'}{(2\pi\i)^2}
 \frac 1{z'(\e)} Q_\g(\e) \left( 2\pi\i \delta(\e-\e') +
 \G^b(\e,\e')Q_\g(\e') \right) \frac 1{z'(\e')} + \ \chi_V'',
 \label{eqn:chi_v}
\end{eqnarray}
where $\chi_V''$ is given by (\ref{eqn:vanvleck2}).

Before concluding this subsection, 
we consider the insulating case.
In this case, apparently $\chi_P = 0$ 
and $\chi_{\rm ins}= \chi_V$ at zero temperature.
By use of the results obtained in the previous sections,
we can show that
\begin{eqnarray}
\chi_{\rm ins} &=& -2g^2 J(J+1) \int \int \frac{d\e d\e'}{(2\pi\i)^2}
 \frac 1{z'(\e)} Q_{\g}^i(\e) \left( 2\pi\i \delta(\e-\e') +
 \G^b(\e,\e')Q_{\g}^i(\e') \right) \frac 1{z'(\e')} + \chi_V''
 \nonumber \\
&=& \ \chi_{\rm ins}^0 \cdot \frac1{z_H(\D_{-}^\ast)} 
 \cdot \frac{z(\D_{-}^\ast)}{z'(\D_{-}^\ast)} \ + \ \ \chi_V'',
 \label{eqn:chi_i}
\end{eqnarray}
where $\chi_{\rm ins}^0= 2g^2 J(J+1) \cdot C\rho(\D_{-}^\ast)$,
which is similar to the unperturbed value given by (\ref{eqn:vanvleck00}).
$\chi_V''$ is given by (\ref{eqn:vanvleck2}).
Thus, we get the simple expression
\begin{eqnarray}
\chi_{\rm ins} \cong \chi_{\rm ins}^0 \cdot \frac1{z_H(\D_{-}^\ast)} .
 \label{eqn:fin-res2}
\end{eqnarray}

From the mathematical point of view, the obtained expressoin for $\chi$,
given by (\ref{eqn:chi}), is identical to (\ref{eqn:total}).
In fact, the energy-integration appearing in the definition of
$\chi$, given by (\ref{eqn:total}),
has been included in the enhancement factor at the Fermi energy, $1/z_H(0)$.
Also, the expression (\ref{eqn:chi_v}) and (\ref{eqn:chi_i}) are 
identical to (\ref{eqn:vanvleck}) mathematically.
However, the expressions obtained in this section show clear physical information :
We can see that 
$\chi$ and $\chi_V$ ( or $\chi_{\rm ins}$ ) are apparently enhanced by the interaction
because they are proportional to the enhancement factor, $1/z(0)$
( or $1/z(\D_{-}^\ast)$ ).

\section{Further Analysis for $\chi$, $\chi_V$ and $\chi_{\rm ins}$}
\subsection{Estimation for the Wilson ratio in the $d=\infty$ limit :
             in case $U\neq0$} 

In the previous section, we get the general expression for 
$\chi$ and $\chi_V$, and find that they are enhanced by $1/z$.
Furthermore, we estimate the Wilson ratio in this section
by examing the magnetic enhancement factor, $1/z_H$.
For this purpose, we consider the $T$-linear coefficient of the specific heat
and the charge susceptibility at first.

(i) \ $T$-linear Coefficient of the Specific Heat \ ; \ $\g$  \\
In the metallic case, the $\g$ is given by,
\cite{Hanzawa_p}
\begin{eqnarray}
\g= \g^0/z_{1}(0) \cong \g^0/z(0),
\end{eqnarray}
where $\g^0$ is the value in case $U=0$.
$\g$ is proportional to the mass-enhancement factor $1/z_1(0) \cong 1/z(0)$.

(ii) \ Charge Susceptibility \ ; \ $\chi_{ch}$  \\
By use of the relation
\begin{eqnarray}
{\left. \frac{\d}{\d \mu} \Si_M(\w) \right| }_{Q_\a=0} = \ddw \Si_M(\w),
\end{eqnarray}
the charge susceptibility in the metallic case is given by
\begin{eqnarray}
\chi_{ch} &=&  \frac{\d \langle n^f \rangle}{\d \mu}
 = \int \depi ( \v_a(\e)+Q_\a(\e) )
 \left( 1- \frac{\d}{\d \mu} \Si_M(\e) \right)  \nonumber \\
&=& \int \depi \v_a(\e)/z_1(\e) \nonumber \\
&+& \int \int \frac{d\e d\e'}{(2\pi\i)^2} \frac 1{z(\e)} Q_\a(\e) \left\{
 2\pi\i \delta(\e-\e')+ 
 \left( \G^{a,\a}(\e,\e')+ (2J+1)T^{a,\a}(\e,\e') \right) Q_\a(\e) \right\} 
 \frac 1{z(\e')},
 \label{eqn:charge}
\end{eqnarray}
where $T^{a,\a}(\e,\e')$ is the antiparallel four-point ( reducible ) 
vertex. ( see Fig. B $\cdot$ 1.)
The first term of (\ref{eqn:charge}) turns out to vanish identically,
as proved by (\ref{eqn:zero}).

Usually, in heavy Fermion systems, $\chi_{ch}$ is considerably suppressed
by the strong Coulomb repulsion between $f$-electrons.
 \cite{Yamada_p}
Here, we assume the following relation approximately :
\begin{eqnarray}
\G^{a,\a}(0,0)= - T^{a,\a}(0,0), 
 \label{eqn:symm}
\end{eqnarray}
which is rigorous for $J=5/2$ single-site Anderson model,
and expected to be valid approximately even for the lattice problem.
( In reality in the periodic system, 
the relation (\ref{eqn:symm}) may be modified by the magnetic correlation
between different sites.
But, we do not consider such an effect now. )
If we put $\chi_{ch} \ll 1$ in the strong coupling limit region, 
we get from (\ref{eqn:charge}) that
\begin{eqnarray}
-\G^{a,\a}(0,0) \cdot Az(0)\rho(0) = \ \frac1{2J} 
 = \ \frac 15  \ \ \ \ \  {\rm for}\ J=5/2.
 \label{eqn:stoner}
\end{eqnarray}

(iii) \ Magnetic Susceptibility \ ; $\chi$ \\
Here, we consider $\chi$ of $J=5/2$ PAM, in case there is no ECF and 
$\rho(\e) \cong$ constant for $0 \le -\e \simle T^\ast$.
At first, we consider the metallic case.
By use of (\ref{eqn:z_H}), (\ref{eqn:stoner}) and 
$\G^{a,\a}(0,0) \cong \G^{b,\b}(0,0)$,
the relation $1/z_H(0) \cong 1.2 \cdot 1/z(0)$ is obtained.
Thus, $\chi$ is given by
\begin{eqnarray}
\chi \sim 1.2 \cdot \chi^0 \cdot \frac 1{z(0)}.
 \label{eqn:R12} 
\end{eqnarray}
So, the obtained Wilson ratio is $\ R \sim 1.2 \cdot R^0$.
( $R^0$ is given by (\ref{eqn:R0}). ) 
In fact, the result (\ref{eqn:R12}) is changed
if the assumption (\ref{eqn:symm}) is incorrect.
Nonetheless, $R \sim 1$ is expected
because the contribution to $R$ from the vertex corrections, 
$T(0,0)$ and $\G(0,0)$, is about $R-R^0 \sim \ 0.2  \ll 1$.
\cite{Storner}
It should be stressed that
the Wilson ratio for $J=5/2$ PAM becomes equal to 
that for $J=5/2$ impurity Anderson model, $R=1.2$,
if the magnetic correlation between different sites can be neglected.
\cite{explain2}
This result is nontrivial in the previous works,
\cite{Zou,Zhang,Nakano}
and is contrastive to that for SU(2) PAM, $R \sim 2$.
\cite{Yamada_p}

On the other hand, because the relation
$0< -\G^{b}(0,0) \cdot C z(0)\rho(0) \simle 1/5$ is expected,
the Van Vleck susceptibility (\ref{eqn:chi_v}) is written by
\begin{eqnarray}
& &\chi_V \sim \chi_V^0 \cdot \frac 1{z(0)},
\end{eqnarray}
where $\chi_V^0 = 2g^2 J(J+1) \cdot C\rho(0)$,
which is similar to the unperturbed value, (\ref{eqn:vanvleck0}).
Thus, $\chi_V$ is enhanced also by $1/z$.

We stress that the RPA-type diagrams are included only in the
polarization factor ( i.e., in $R/R^0$ ),
and never included in the enhancement factor $1/z$ or $1/z'$
which brings the highly enhanced 
magnetic susceptibility observed in the heavy Fermion systems.
( see Appendix E. )
This situation also holds in the SU(N)-PAM.


In conclusion, the Van Vleck susceptibility for the $J=5/2$ PAM
is strongly enhanced by $1/z$,
both in the metallic case and in the insulating case,
and $\chi_{\rm ins}$ for the $J=5/2$ PAM 
is much larger than that in the insulating SU(N)-PAM without ECF.

\subsection{Consideration on the $d=3$ system : in case $U\neq0$}

Here, we briefly consider
the magnetic susceptibility in the three dimensional system
beyond the $d=\infty$ approximation.
The relation $\chi= \chi^0 \cdot (1/z_H)(z/z')$ is also derived in $d=3$,
because the equation (\ref{eqn:important})
is also satisfied in $d=3$
by the replacement of $Q_\gamma(\w)$ with 
$Q_\gamma(\w) \cdot z(0)\delta(E_k^\ast)$ in (\ref{eqn:abQ}).
On the contrary, 
the relation $1/z \sim 1/z'$ is not guaranteed in $d=3$,
because the relation (\ref{eqn:diag}) is rigorous 
only for the $d=\infty$ case.

In the $d=3$ system,
there exist the correction from the non-local part of the 
irreducible four-point vertex, $\G_{\rm nl}^I$,
which violates the relation (\ref{eqn:diag}).
Fortunately, SOPT for $J=5/2$ PAM
by means of the $1/d$-expansion 
shows that the correction term $\G_{\rm nl}^I$ 
is approximately negligible in $d=3$, 
so (\ref{eqn:diag}) will be satisfied very well.
As a result, the relation 
$\chi_V/\chi_V^0 \sim \chi_P/\chi_P^0 \sim 1/z$ 
holds also in the three dimensional $J=5/2$ PAM.

At last, we make a more general but qualitative consideration 
on the original definition for $\chi_V$ given by (\ref{eqn:vanvleck}).
Approximately, we can transform (\ref{eqn:vanvleck}) to 
$\chi_V \sim -\int^0_{-\infty} d\e \rho(\e) \dde (\L^\w(\e)z(\e))$,
where $\rho(\e)$ is the DOS for the $f$-electrons.
( see (\ref{eqn:test1}) or (\ref{eqn:test2}). )
Here, we consider the quasiparticle contribution to $\chi_V$, 
i.e., the contribution from the integration range, $0<-\e\simle T^\ast$.
It is not enhanced 
only when the relation $|{\hat \L}^\w(\e) \cdot z(\e)| \cong 1$ 
is satisfied.
The enhancement of $\chi_V$ is not determined by the value of the
enhancement factor at the Fermi energy,
${\hat \L}^\w(0)$, but by the $\e$-dependence of ${\hat \L}^\w(\e)$.
( From the definition of $T^\ast$,
$|z(0)/z(-T^\ast)| \ll 1$ should be satisfied, and $z(\pm \infty)=0$. )
From the result of this paper, $\chi_V/\chi_V^0 \sim 1/z$ 
in $d=\infty$ $J=5/2$ PAM,
$|{\hat \L}^\w(-T^\ast)|/|{\hat \L}^\w(0)| \simle O(1)$ should be realized.
Such a property of ${\hat \L}^\w(\e)$ 
will not depend so much on the dimension of the system.
Thus, $\chi_V$ should be enhanced
by the strong correlation in three-dimensional systems.

\section{Discussions}
Here, we summarise the conclusions of this paper.
At first, we obtain the general but abstract
expression for the magnetic susceptibility 
on the basis of the orbitally degenerate FLT.
In the next stage, we employ two simplifications :
One of them is to assume no ECF and the spherical Fermi surface,
and another is the $d=\infty$ approximation.
Since our aim of this paper is to elucidate unambiguously 
the essential properties 
of the susceptibility under the influence of the strong Coulomb interaction,
these over-simplifications will be allowed.
Needless to say, it is significant to confirm the property
for {\it the zero ECF limit case}.
The opposite limit case, where only the lowest Kramers doublet
contributes to the ground state, has already been discussed elsewhere.
\cite{Hanzawa_p,Yamada_v} ( see Appendix A. )

After the two simplifications,
we make further analysis on the expressions both for $\chi$ and for $\chi_V$,
taking account of all the vertex corrections in a consistent way.
Below, we summalize the results of this paper,
which will hold qualitatively even in the three-dimensional case.

(i) Metallic Case \  \\
In this case the Fermi energy lies below the hybridization gap,
which is a prototype of the ( Ce-compound ) heavy Fermion systems.
 \cite{insulator}
The total susceptibility is given by 
$\chi= \ \lim_{k\rightarrow0}\chi_k(0) = \ \chi_V+\chi_P$.
In $d=\infty$ approximation, 
we get the simple expression,
$\chi= \chi^0 / z_H(0)$.
( $1/z_H(0)$ is the magnetic enhancement factor at the Fermi energy. )
We can also express $\chi$ by use of the mass enhancement factor
$1/z(0)$ as
$\ \chi \sim 1.2 \cdot \chi^0 / z(0)$,
which means that the Wilson ratio is $R \sim 1.2$.
In the same way, the Van-Vleck susceptibility is also expressed as
$\ \chi_V \sim 1.2 \cdot \chi_V^0 / z(0)$.
In conclusion, both $\chi$ and $\chi_V$ are proportional 
to the mass-enhancement factor.
Our conclusion contradicts to the conclusion in ref. \cite{Zou}.

(ii) Insulating Case \  \\
In this case the Fermi energy lies in the hybridization gap.
This is a prototype of the {\it so called} Kondo insulators,
some of which exhibit the large magnetic susceptibility experimentally
at $T=0$.
\cite{insulator}
In this case, $\chi_P=0$
and the magnetic susceptibility $\chi_{\rm ins}$ is given only by the 
Van Vleck susceptibility at zero temperature.
we get the simple expression as
$\ \chi_{\rm ins}= \chi_{\rm ins}^0 / z_H(\D_{-}^\ast)$,
where $\D_{-}^\ast$ is the renormalized lower hybridization edge.
Because $1/z_H(\D_{-}^\ast) \cong 1/z_H(0)$ is expected,
the magnetic susceptibility for the orbitally degenerate model 
is strongly enhanced.
This result is consistent with the result (i) because 
the Van Vleck susceptibility will be insensitive to
the state of the Fermi surface.
So, $\chi_V$ will be little affected 
by the superconducting transition.

As our calculations are very lengthy and involved, we briefly summarise the
mathematical analysis for the Van Vleck susceptibility, $\chi_V$ :
As is shown by (\ref{eqn:def_v}),
$\chi_V$ is given by the $\w$-limit of the dynamical magnetic susceptibility.
After the $d=\infty$ approximation,
it is written as
$\chi_V \sim \v_b ( 1+\G^b \v_b)$, where $\v_b$ and $\G^b$ are
given by (\ref{eqn:ddw_green}) and (\ref{eqn:BS1}), respectively.
( Here, the symbols for the energy-integration are implicit. )
On the other hand, $1/z(\w) \sim ( 1+\G^a \v_a )$.
$\G^a$ and $\G^b$ are related by the Bethe-Salpeter equation given by 
(\ref{eqn:BSab1}), $\G^b= \G^a+ \G^a Q_\gamma\G^b$.
( Here, we have identified $\G^a$ with $\G^{a'}$ for simplicity. )
By use of this equation, we can finally show that 
$\chi_V \sim \int d\w \{ 1/z(\w) \cdot Q_\gamma(\w) \cdot 1/z(\w)\} \ \sim \ 1/z(0)$.
In the same way, $\chi \sim 1/z(0)$ is derived.

Here, we make the physical consideration on $\chi_P$ and $\chi_V$
in terms of the one-body picture,
which was done previously by Anderson and Zou.
\cite{Anderson}
In the one-body picture, $\chi_P/\chi_P^0 \sim 1/z(0)$
and $\chi_V/\chi_V^0 \sim ( {E^f}-\mu )/( {E^f}^\ast-\mu)$,
where $z(0)$ is the renormalization factor at the Fermi energy
and ${E^f}^\ast$ is the renormalized $f$-electron spectrum.
In the mean-field approximation, where the 
frequency-independent renormalization factor $z_{\rm const}$ is assumed,
then ${E^f}^\ast$ is strongly renormalized
towards the Fermi energy so that $\chi_V/\chi_V^0 \sim 1/z_{\rm const}$.
On the other hand, from the viewpoint of the FLT,
renormalization is caused by the strong energy-dependence of the selfenergy.
The renormalized value of ${E^f}^\ast$ critically depends on
the energy range of the coherent region ( around the Fermi energy ),
in where the energy dependence of the selfenergy is large.
Thus, the question '{\it to what extent $\chi_V$ is enhanced or not
by the strong correlation }'
is never trivial within the one-body picture.
The simple mean-field approximation never answer this question.

Our analysis shows that the excitation of the quasiparticles,
which are well defined only within $T^\ast$,
bring the enhancement of $\chi_V$, $\chi_V/\chi_V^0\sim 1/z_H(0)$.
Our work also suggests that $E^f$ is strongly renormalized
to the Fermi energy.

Finally, we point out some future problems.
At first, 
it is interesting to estimate the influence of the electronic crystal field,
or the shape of the Fermi surface ( i.e., the shape of the lattice ).
These effects are ignored in this paper.
Secondly, 
the effect of the antiferromagnetic fluctuation should be taken into
account correctly.
The antiferromagnetic fluctuations 
will make the magnetic susceptibility $\chi$ smaller than our prediction,
although they cannot change the value of $\chi$ and $\chi_V$ drastically.
In fact, some heavy Fermion compounds
are under the influence of the prominent antiferromagnetic fluctuations,
which may be the driving force to the superconducting state.
Such a study will give us much information on the
electronic properties of the heavy Fermion systems.

\vskip 1.5cm

\section*{Acknowledgements}
The authors would like to K. Ueda for valuable discussions and comments.
They are also grateful to F. C. Zhang for useful discussions,
and to T. Mutou for informing us some experimental data
on Kondo insulators.
This work is supported by Grant-in-Aid for Scientific Research
from the Ministry of Education, Science and Culture of Japan.

\vskip 2.0cm

\appendix
\section{Theory of the Magnetic Susceptibility for SU(N)-PAM}  
In this Appendix, we briefly consider the magnetic susceptibility for
SU(N)-PAM.
In case $U=0$, it is easy to show that
$\chi_V^0$ is much smaller than $\chi_P^0$.
Here, we study the case $U \neq 0$, 
making no use of the $d=\infty$ approximation.
The Hamiltonian without ECF is given by
\begin{eqnarray}
& &H_0 = \sum_{\k M} \epsilon_{\k} c_{\k M}^\dagger c_{\k M}
       + \sum_{\k M} E^f f_{\k M}^\dagger f_{\k M}
       + \sum_{M\k} ( V f_{\k M}^\dagger c_{\k M}
         + V c_{\k M}^\dagger f_{\k M} )  ,
 \label{eqn:a1}  \\
& &H_1 = \frac U2 \sum_{\k \k ' \q M\neq M'}
 f_{\k - \q M}^\dagger f_{\k '+ \q M'}^\dagger f_{\k 'M'} f_{\k M} .
 \label{eqn:a2}
\end{eqnarray}
We note that the conduction electrons have the six-fold degeneracy.
This model is analyzed usually by the slave boson technique.
\cite{SUN}

Here, we assume the existence of the magnetic field $H$ along $z$-axis.
At first, we assume that 
both $E^f$ and $\e_{\k}$ appearing in (\ref{eqn:a1}) are
shifted by the same Zeeman energy as
\begin{eqnarray}
\left\{
\begin{array}{l}
E_{M}^f =E^f + g_f \mu_{\rm B} M \cdot H,    \\
\e_{\k M}= \e_{\k} + g_c \mu_{\rm B} M \cdot H.
 \label{eqn:a10}
\end{array}
\right.
\end{eqnarray}
Only in the special case $g_f= g_c$, 
the magnetization operator ${\hat M}$ is conserved
and $\chi$ is derived after the Luttinger's manner.
\cite{Luttinger,Yamada_p}
But, we consider the case $g_f \neq g_c$ here.
The Green's functions for $f$-electrons $G_{kM}(\w)$ 
and that for the conduction electrons $G_{kM}^c(\w)$ are given by
\begin{eqnarray}
G_{\k M}(\w) &=&
 \left(\w+\mu-E^f -\Si_{\k M}(\w)- \frac{V^2}{(\w+\mu-\e_{\k})^2}\right)^{-1}, \\
G_{\k M}^c(\w) &=& G_{\k M}(\w)\cdot V^2/(\w-\mu-\e_\k)^2,
\end{eqnarray}
where $\Si_{kM}(\w)$ is the selfenergy.

Here, we define the three types of the enhancement factors.
\begin{eqnarray}
& &\frac 1{z(\w)} = 1- \ddw \Si_{k M}(\w), \\
& &\frac 1{z_H(\w)} = 1+ \frac1{g_f M} \ddH \Si_{k M}(\w), \\
& &\frac 1{z'(\w)} = \left. \frac 1{z_H(\w)} \right|_{\w{\mbox -}{\rm limit}},
 \label{eqn:a50}
\end{eqnarray}
where (\ref{eqn:a50}) is defined by (\ref{eqn:Lam_k}).
From the definition of $\chi_P$, given by (\ref{eqn:pauli}), we get
\begin{eqnarray}
& &\chi_P = {\chi_P^f}^0 \cdot \left(\frac 1{z'(0)}+ \frac{g_c}{g_f}
 \frac{V^2}{(\mu-\e_{k_F})^2} \right) \cdot  
\left(\frac 1{z(0)}+ \frac{V^2}{(\mu-\e_{k_F})^2} \right)^{-1}
\cdot \left( \frac 1{z_H(0)}+
 \frac{g_c}{g_f} \frac{V^2}{(\mu-\e_{k_F})^2} \right),
 \label{eqn:a31} \\
& &\ \ \ \ {\chi_P^f}^0= \sum_M g_f^2 M^2 \rho(0), \nonumber
\end{eqnarray}
where $\rho(0)$ is the DOS of the $f$-electrons at the Fermi energy.
On the other hand, by use of (\ref{eqn:vanvleck}), we get
\begin{eqnarray}
\chi_V &=& - \frac 1N \sum_{\k M} g_f^2 M^2 
 \int_{-\infty}^\infty \dwpi \cdot \left(
 \frac 1{z'(\w)} + \frac{g_c}{g_f} \frac{V^2}{(\w+\mu-\e_{\k})^2}
 \right) / \left( \frac 1{z(\w)} + \frac{V^2}{(\w+\mu-\e_{\k})^2}
 \right) \nonumber \\ 
& &\times \ddw \left( G_{\k M}(\w)+
 \frac{g_c}{g_f} G_{\k M}^c(\w) \right).
 \label{eqn:a30}
\end{eqnarray}

In case $g_f= g_c$, $1/z(\w)=1/z'(\w)$ is satisfied rigorously.  Then,
$\chi_P= \chi_P^0 /z_H(0)$ and $\chi_V=0$ is derived.  
Although $1/z(\w)=1/z'(\w)$ is not satisfied rigorously
in case $g_f \neq g_c$,
its discrepancy will be of order $\sim (V/D)^2/z$ at
most because the relation $\frac{\d}{\d (g_fH)} \Si_M(0)$ $\sim$
$\frac{\d}{\d (g_cH)} \Si_M(0) \cdot D^2/V^2$ is expected from
(\ref{eqn:ddw_S}) and ref. \cite{explain}.

Thus, $\chi_V$ is negligible compared with $\chi_P$.
So, the susceptibility for the insulating case, $\chi_{\rm ins}$, 
cannot become large.
This reasoning has been already applied to the system with the Kramers
doublet ground state in the strong ECF limit.
\cite{Hanzawa_p,Yamada_v}
In the Gutzwiller approximation, $\chi_V/\chi_V^0$ in this model
is not enhanced because it is proportional
to the conduction electron DOS, which is not renormalized.
But the more detailed analysis should be required.
Finally, we point out that $\chi_V=0$ for any model where $\hat M$ 
is conserved,
because $1/z(\w)= 1/z_H(\w)$ is satisfied then.

\section{Some Properties of the Irreducible Vertex} 
At first, we show that we can ignore the Pauli's principle in
(\ref{eqn:hint2}).  The term excluded in (\ref{eqn:hint2}) due to the
Pauli's principle
is, in the real-space representation, given by
\begin{eqnarray}
  U \sum_{ \{ i \},M } f_{iM}^\dagger f_{iM} f_{iM}^\dagger f_{iM},
 \label{eqn:h} 
\end{eqnarray}
where $\{ i \}$ represents the set of the $f$-electron sites.
Apparently, (\ref{eqn:h}) represents a constant energy shift and is
independent of $M$ in the paramagnetic state.  Therefore, we can
ignore the Pauli's principle in the following discussion if all the
diagrams are taken into consideration.  In other words, contributions
from the diagrams violating the Pauli's principle cancel out in each
order of the perturbation in the paramagnetic state.  \cite{Yosi}

Then, we investigate the general properties of four-point vertices.
We classify four-point vertices into the
parallel vertex $\G$ and the anti-parallel vertex $T$, as is shown by
Fig. B $\cdot$ 1.  Needless to say, $T$ never contributes to $\ddw
\Si_M(\w)$.  Moreover, $\ddH \Si_M(\w)$ also have no contribution from
$T$ because Pauli's principle is ignored now.  

In $d=\infty$ limit, by neglecting the Pauli's principle, we can regard that the
irreducible parallel four-point
vertex $\G^I$ satisfy the relation
\begin{eqnarray}
\G_{MM,M'M'}^I(\e,\e') = \G^I(\e,\e') \cdot \delta_{MM'},
 \label{eqn:Bv}
\end{eqnarray}
because the local Green's function is independent of $M$ and diagonal 
with respect to $M$.

On the other hand, when we investigate $\frac{\d}{\d \mu} \Si_M(\w)$,
we have to take account of the contribution from both $T$ and $\G$.
The irreducible antiparallel four-point vertex $T^I$ satisfy
the relation
\begin{eqnarray}
T_{MM,M'M'}^I(\e,\e') = T^I(\e,\e') \ \ \ \ \ \ \ {\rm for \ any}
  \ \ \ M, M',
\end{eqnarray}
by neglecting the Pauli's principle.

\section{The Consideration on $\v_{a(b)}(\w)$ and
 the Proof of (66) and (69)} 
At first, we consider the following calculation
as a preparation of calculating the enhancement factor.
\begin{eqnarray}
\int_{-\infty}^{\infty} \dwpi \G'(0,\w) \v_{a(b)}(\w) \G'(\w,0),
 \label{eqn:example}
\end{eqnarray}
where $\G'(\e,\e')$ is an irreducible four-point vertex with respect to
the Coulomb interaction $U$.
Then,
\begin{eqnarray}
\{ { \mbox { eq. (\ref{eqn:example})} } \} &=& 
 \int_{C} \dwpi \G'(0,\w) \v_{a(b)}(\w) \G'(\w,0) \nonumber \\
&=& \int_{-\infty}^0 \frac{d\zeta}{\pi} \left\{ 
 {\rm Im} \{ \v_{a(b)}(\zeta) \} {\rm Re} \{ {\G'}^2(\zeta,0) \}
 + {\rm Re} \{ \v_{a(b)}(\zeta) \} {\rm Im} \{ {\G'}^2(\zeta,0) \} \right\}.
 \label{eqn:example2}
\end{eqnarray}
In the first line of the r.h.s. of (\ref{eqn:example2}),
we have changed the path of $\w$-integration as shown in Fig. C$\cdot$1.
Even if $\G'$ is replaced with reducible vertex $\G$,
this procedure is correct
because $\v_a$ $\sim$ $\v_b$ $\sim$ $\w^{-2}$ for $\w \simge D$.
The last term in (\ref{eqn:example2}) can be obtained by substituting
the following spectrum representations
( which is possible because $\v_{a(b)}(\pm\infty) = {\G'}^2(\pm\infty,0)= 0$.) ,
\begin{eqnarray}
& &\v_{a(b)}(\w)= -\int \frac{d \zeta}{\pi} \left\{
 \frac{\theta(\zeta)}{\w-\zeta+ \i \eta}
 + \frac{\theta(-\zeta)}{\w-\zeta- \i \eta} \right\} \cdot
 {\rm Im} \v_{a(b)}^R(\zeta), 
 \label{eqn:spectrum1} \\
& &{\G'}^2(\w,0)= -\int \frac{d \zeta}{\pi} \left\{
 \frac{\theta(\zeta)}{\w-\zeta+ \i \eta}
 + \frac{\theta(-\zeta)}{\w-\zeta- \i \eta} \right\} \cdot
 {\rm Im} {{\G'}^R}^2 (\zeta,0),
 \label{eqn:spectrum2}
\end{eqnarray}
and by performing $\w$-integration at first.
In the last line of (\ref{eqn:example2}), 
only Re$\{ {\G'(\zeta,0)}^2\}$
for $0< -\zeta \simle T^\ast$ contributes predominantly to the 
value of the $\zeta$-integration, which is largely enhanced.
Thus, in heavy Fermion systems, the behavior of
${\rm Im}\v_{a(b)} (\e)$ only for $0 \leq -\e \simle T^\ast$
is dominant in determining the enhancement factor.

In the next stage, we prove the important relation given by
(\ref{eqn:important}).  
The spectral representations for the local Green's functions,
given by (\ref{eqn:lgreen}), are
\begin{eqnarray}
  g^{(f)}(\w) &=& \int {d \zeta} \left\{
  \frac{\theta(\zeta)}{\w-\zeta+ \i \eta} +
  \frac{\theta(-\zeta)}{\w-\zeta- \i \eta} \right\} \cdot
  \rho^{(f)}(\zeta),
 \label{eqn:b1} \\
\rho(\zeta)&=& -\frac1\pi {\rm Im} \frac 1N \sum_\k G_{\k MM}^R(\zeta) =
 -\frac1\pi {\rm Im} g^R(\zeta), \nonumber \\ 
\rho^f(\zeta)&=&
 -\frac1\pi {\rm Im} \frac 1N \sum_\k d_{\k MM} {G_{\k}^f}^R(\zeta),
 \nonumber
\end{eqnarray}
where $\rho^{f}(\zeta)$ represents the DOS of the localized
$f$-electrons, and $d_{kMM}$ is given by (\ref{eqn:green2}).
Apparently, $\rho(\zeta) \geq \rho^f(\zeta)$ for any $\zeta$ and
$\rho^f(0)=0$.  On the other hand, we can also express the
particle-hole Green's functions
in the spectral representation as follows ( see (\ref{eqn:b1}) ) :
\begin{eqnarray}
  & &\frac 1N \sum_\k {G_\k(\w)}^2 = - z(\w) \cdot \frac 1N \sum_\k \ddw
  G_\k(\w) \nonumber \\ & &\ = -z(\w) \int d \zeta \left\{
  \frac{\theta(\zeta)}{\w-\zeta+ \i \eta} +
  \frac{\theta(-\zeta)}{\w-\zeta- \i \eta} \right\} \cdot \frac{\d}{\d
    \zeta} \rho(\zeta) + 2 \pi \i z(0) \rho(0) \cdot \delta(\w),
 \label{eqn:C2}
\end{eqnarray}
where in deriving the last line above, we have done the partial
integration.

At first, we investigate the case where the system is metallic at
$T=0$, i.e., $\rho(0) \neq 0$. ( see Fig. 1(a).)
For $0<-\w \simle T^\ast$,
\begin{eqnarray}
  & &\rho(\w) \cong \frac 1N \sum_\k a_\k(\w) \delta(\w-E_\k^\ast) \cdot
  \frac{3V^2}{{( \w+\mu-E^f-\Si(\w))}^2} \nonumber \\ & &\ \ \ =
  \rho_c(\w) \cdot \frac{3V^2}{{( \w+\mu-E^f-\Si(\w))}^2}.
\end{eqnarray}
So, we obtain
\begin{eqnarray}
  \ddw \rho(\w) \cong -2 \rho(\w) \cdot \frac1{\w+\mu-E^f-\Si(\w)}
  \cdot \frac1{z(\w)} + \ddw {\rho_c}(\w) \cdot \frac{3V^2}{{(
      \w+\mu-E^f-\Si(\w))}^2}.
 \label{eqn:ddw_rho}
\end{eqnarray}
Considering that the first term of the r.h.s. of (\ref{eqn:ddw_rho})
has the enhancement factor $1/z$ contrary to the last term, we can
neglect the last term of (\ref{eqn:ddw_rho}) for $0 \leq -\w \simle
T^\ast$ in metallic heavy Fermion systems.  On the other hand,
$\rho^f(\w)$ is non-zero only for $\w \sim |{E^f}^\ast - \mu| >0$,
corresponding to the renormalized $E^f$ spectrum, and $\w \sim \pm U/2
, \ \ ( U \gg T^\ast )$, corresponding to the broad satellite on both
sides of the Fermi energy \cite{paper3}.  Especially, $\rho^f(0)= 0$.
Thus, both Im$G_k^f(\w)$ and Im${G_k^f}^2(\w)$ make
little contribution to the behavior of (\ref{eqn:important}) or
(\ref{eqn:important2}), so they are negligible.
By use of the relation for $0< -\w \simle T^\ast$,
\begin{eqnarray}
  {\rm Im}\frac 1N \sum_\k (-1/\pi)G_\k^R(\w)\cdot
  \frac{\w+\mu-\e_\k}{3V^2} \cong \frac1{\w+\mu-E^f-\Si(\w)} \cdot
  \rho(\w),
\end{eqnarray}
we can show from (\ref{eqn:ddh_green}) that
\begin{eqnarray}
& &{\rm Im}\v_b(\w) \cong {\rm Im} \left\{ B \frac 1N \sum_\k
  {G_\k}^2(\w) + (A-B)\left( \frac 1N \sum_\k G_\k^2(\w) - 2\pi \i
  z(0)\rho(0) \delta(\w) \right) \right\} \nonumber \\ & &\ \ \ \ \ \ 
  \ \ \ \ \ \cong {\rm Im} \{ \v_a(\w) + Q_\g(\w) \}, \nonumber
\end{eqnarray}
for $0< -\w \simle T^\ast$.  Here we have used (\ref{eqn:C2}) and
(\ref{eqn:ddw_rho}).  We stress that when $U=0$ and $\rho_c^0(\w)=
{\rm constant}$ with respect to $\w$,
the relation (\ref{eqn:important}) is rigorous for $\w \leq 0$.

In the same way, we consider the case where the system is insulating,
i.e., $\rho(0)=0$. ( see Fig. 1(b). ) We prove the important relation
given by (\ref{eqn:important2}).  Note that $\rho(\w)=0$ for
$\D_{-}^\ast < \w < {E^f}^\ast-\mu$, where $\D_{-}^\ast$ and
${E^f}^\ast$ are the renormalized lower edge of the hybridization gap
and the renormalized local $f$-electron level, respectively.  They are
given by $\D_{-}^\ast \cong \D_{-} \cdot z(0)$ and ${E^f}^\ast \cong (
E^f-\mu ) \cdot z(0) +\mu$. \cite{paper3}  In this insulating case,
taking account of the fact ${\rm Im}\Si(\w) =0$ for $\D_{-}^\ast \leq
\w \leq E^{f\ast}-\mu$,
we can show that
\begin{eqnarray}
  \rho(\w) \cong \rho_c(\w) \frac{3V^2}{(\w+\mu-E^f-\Si(\w))^2} \cdot
  \theta(\D_{-}^\ast-\w),
 \label{eqn:c11}
\end{eqnarray}
for $\D_{-}^\ast-T^\ast \simle \w < E^{f\ast}-\mu$.
So, we get for $\w \leq 0$,
\begin{eqnarray}
  \ddw \rho(\w) \cong \{ {\mbox {\rm eq.(\ref{eqn:ddw_rho})}} \} -
  \rho(\D_{-}^\ast) \delta(\w-\D_{-}^\ast).
 \label{eqn:c10}
\end{eqnarray}
Thus, for $0 \simle -\w \simle -\D_{-}^\ast + T^\ast$, we can show the
relation (\ref{eqn:important2}) in the insulating case.  Here, we
comment that if the anisotropy of the Brillouin zone is
taken into account in $d=3$ system, the step function in
(\ref{eqn:c11}) becomes a continuous function
because of the van Hove singularity.
So, the delta function
in (\ref{eqn:c10}) has finite ( but narrow ) width.  In this sense, the
relation (\ref{eqn:important2}) is less universal than the relation
(\ref{eqn:important}).

\section
{Numerical Calculations for $1/z(\w)$ and $1/z'(\w)$} 

In this Appendix, we explain the method and the results of the
numerical calculation by SOPT and SC-SOPT with respect to $U$ in the
$d=\infty$ limit.
We calculate both $1/z(\w)$ and $1/z'(\w)$ 
and check the relation (\ref{eqn:propor}),
which is derived in the analytical way.
The relation (\ref{eqn:propor}) is very significant
because our main results
(\ref{eqn:fin-res}) and (\ref{eqn:fin-res2}) are based on it.


At first, we calculate $ 1/z(\w) $ and $ 1/z'(\w) $ by SOPT, which are
depicted in Fig. D$\cdot$1(a) and D$\cdot$1(b), respectively.  Here,
we use the constant DOS for the conduction electrons, i.e.,
$\rho_c^0(\w) = \rho_c^0$.  The numerical results of their real part
are shown in Fig. D$\cdot$2.  The obtained $T^\ast$ in SOPT is about
$|E^f-\mu|$.  Surprisingly, the resultant $ 1/z(\w)$ and $ 1/z'(\w)$
are almost the same not only for $|\w| \simle T^\ast$ but also for
$|\w| \gg T^\ast$.  
But, this result is modified when $\rho_c^0(\w)$ has drastic energy
dependence for $0 \leq -\w \simle T^\ast$.  By use of SOPT, we also
calculate $ 1/z(\w)$ and $ 1/z'(\w)$ for insulating case.  We put
Fermi energy $\mu$ in the middle of $\D_{-}$ and $E^f$.  The numerical
results of their real parts are shown in Fig. D$\cdot$3.  In the
framework of the SOPT, $\D_{-}$ is renormalized to be $\D_{-}^\ast
\cong z(0) \cdot \D_{-}$.

In the second stage, we calculate $ 1/z(\w)$ and $\ 1/z'(\w)$ by
SC-SOPT.  
The irreducible four-point vertex $\G^I$ is composed of at most two
local Green's functions, and is shown by Fig. D$\cdot$4.  In SC-SOPT,
we construct $ 1/z(\w)$ and $ 1/z'(\w)$ with $\G^I$ and $g(\w)$ by
referring to their definitions, (\ref{eqn:ddw_S}) and
(\ref{eqn:ddh_c}), respectively.  ( Then, $U$-linear terms of $\G^I$
turn out to give no contribution, as is shown below. ) The numerical
results of their real parts are shown in Fig. D$\cdot$5.  Here, we
assume $\rho_c^0(\w) \propto ( \w+D )^2 $.
The obtained $T^\ast$ is renormalized to be 
$\sim z(0) \cdot |E^f+\Si(0)-\mu|$.

Surprisingly, in spite of $\rho_c^0(\w) \neq \rho_c^0$, the resultant
$ 1/z(\w)$ and $ 1/z(\w)$ are almost the same not only for $|\w|
\simle T^\ast$ but also for $|\w| \gg T^\ast$.  The discrepancy
becomes almost invisible in case $\rho_c^0(\w)= \rho_c^0$.
It becomes large only when $\rho_c^0(\w)$ has 
drastic energy-dependence for $0 \leq -\w \simle T^\ast$.  
In real heavy Fermion systems, $T^\ast$ is
renormalized as is pointed out by SC-SOPT, so the susceptibility will
be quite insensitive to the energy dependence of the $\rho_c^0(\w)$.

In calculating for $1/z'(\w)$ in case $\rho_c^0(\w)\neq\rho_c^0$, we
do not fix the value of $C$ to (\ref{eqn:c_a_b}), but adjust it so as
to the equation (\ref{eqn:zero}) is satisfied.  Then, the obtained $
1/z'(\w)$ has no RPA-type enhancement, as well as $1/z(\w)$.  The
resultant values of $C_{\rm cal}$ are given by Table D $\cdot$ I.  The
larger the $1/z(0)$ is, the smaller $|C-C_{\rm cal}|$ becomes, as is
expected in Appendix C. ( see (\ref{eqn:ddw_rho}). ) Taking account of
the smallness of $1/z$ in our calculation, we can say that our
numerical result supports (\ref{eqn:propor}) quite well
even in case $\rho_c^0(\w) \neq \rho_c^0$.

\vskip 1cm
\hskip 3cm 
\begin{tabular}{|c|c|c|} \hline
 U & $C_{\rm cal}$ & $1/z(0)-1$ \\ \hline \hline
 \ 0.5 \ & \ 0.143 \ & 4.67 \\ \hline
 0.3 & 0.140 & 3.64 \\ \hline
\end{tabular}

\vskip 0.5cm

\noindent
Table D$\cdot$I \ : \ 
The values of $C_{\rm cal}$ which
    satisfies by eq. (\ref{eqn:zero}) by SC-SOPT.  All the parameters
    except for $U$ are the same as that used in Fig. D$\cdot$5.

\vskip 1.5cm

Here, we do not calculate the magnetic enhancement factor $ 1/z_H(\w)=
1+ \ddHM \Si_M(\w)$ because $ 1/z_H(\w)$ calculated by SC-SOPT
will contain much contribution from RPA-type diagrams.  We know that
their contribution is limited only to the polarization factor, $R/R^0
\sim 1.2$.  ( see \S 6 and Appendix E. ) On the calculation of $
1/z_H(\w)$ by SC-SOPT, we have to take caution against the unphysical
enhancement of the Wilson ratio caused by the RPA-type diagrams.  In
the rigorous perturbation calculation, most of them turn out to be
canceled with other diagrams not contained in SC-SOPT, and amount to a
little contributions totally.

\section
{Proof of (79)} 
In this appendix, we examine the relation $\chi_V''=0$, shown by
(\ref{eqn:zero}), in detail.
To do this, we study the following integrals at first :
\begin{eqnarray}
  \frac 1N \sum_\k \int_{-\infty}^{\infty} \depi G_\k^2 (\e)/z'(\e) &=&
  \frac 1N \sum_\k (-2\i){\rm Im} \int_{-\infty}^{0} \depi G_\k (\e)
  \cdot \dde \{ z_1(\e)/z'(\e) \},
 \label{eqn:test1} \\
 \frac 1N \sum_\k \int_{-\infty}^{\infty} \depi {G_\k^f}^2 (\e)/z'(\e)
 &=& \frac 1N \sum_\k (-2\i){\rm Im} \int_{-\infty}^{0} \depi G_\k^f
 (\e) \cdot \dde \{ z(\e)/z'(\e) \},
 \label{eqn:test2}
\end{eqnarray}
where we have done partial integration.  
Because $\dde (z_{(1)}(\e)/z'(\e)) \ll 1$ for $0<-\e \simle T^\ast$
is expected, both
(\ref{eqn:test1}) and (\ref{eqn:test2}) are not enhanced.  As
$\chi_V''$ is expressed by both (\ref{eqn:test1}) and
(\ref{eqn:test2}) ( see (\ref{eqn:important}) and
(\ref{eqn:vanvleck2}) ), we conclude that $\chi_V''$ is not enhanced,
and small quantity.  Even if $\chi_V'' \neq 0$, a slight modification
of the value of $C$ in the definition of $Q_\gamma(\w)$ can make
$\chi_V'' =0$.  ( As is shown in Appendix D, this is confirmed well
within SC-SOPT. ) This reason is as follows : the behavior of
$z'(\e)$ for $0<-\e\simle T^\ast$ is sensitive to the value of $C$,
and the absolute value of (\ref{eqn:test1}) is largely enhanced
when (\ref{eqn:propor}) is not satisfied.  
Thus, the relation
(\ref{eqn:zero}) is justified.

$1/z'(\w)$ can be rewritten as
\begin{eqnarray}
  1/z'(\w)= \left( 1+ \int \depi \G_{\rm non{\mbox-}tad}^{a'}(\w,\e)
  \v_{a'}(\e) \right) \cdot \left( 1+ U\chi_V''(3/2g^2 J(J+1))
\right),
\end{eqnarray}
where $\G_{\rm non{\mbox-}tad}^{a'}(\w,\e)$ is derived from
$\G^{a'}(\w,\e)$ by dropping the reducible terms with respect to $U$.
When we choose the value of $C$ so that $\chi_V''=0$ is satisfied, (
which will be close to the value given by (\ref{eqn:c_a_b}), ) the
enhancement factor $1/z'(0)$ has no contributions from any irreducible
vertices with respect to $U$, i.e., from any tadpole
diagrams. ( see Fig. E$\cdot$1. )
Moreover, the mass-enhancement factor $1/z(0)$
also has nothing to do with any tadpole diagrams, that is,
\begin{eqnarray}
  1/z(\w)=
  1+ \int \depi \G_{\rm non{\mbox-}tad}^{a}(\w,\e) \v_{a}(\e).
\end{eqnarray}
Then, we can conclude that both $1/z(\w)$ and $1/z'(\w)$ has
no contributions from RPA-type enhancement.  This results suggests
that the relation $1/z(\w) \cong 1/z'(\w)$ holds for a wider range of
$\w$.



\vskip 2cm

\noindent {\bf Figure}

\begin{itemize} 
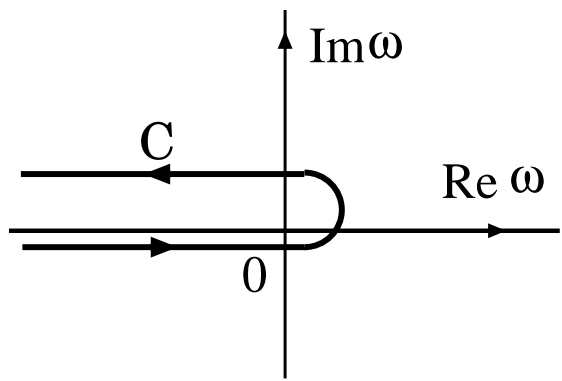
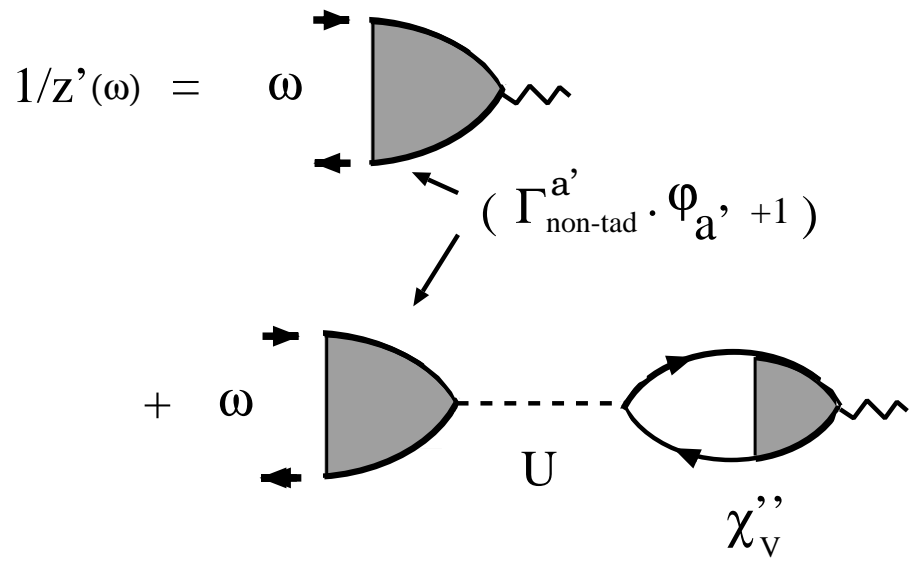
\item{ Fig. 1 \ : \ \ \ 
 Effective band structure for $J=5/2$ PAM.
 $E^f$ is the four-fold degenerate local $f$-electron spectrum, and
 $E_k^\ast$ is the two-fold degenerate quasiparticle spectrum,
 respectively.  (a) Fermi energy $\mu$ lies below $\D_{-}+\mu$. The
 system is metallic.  (b) $\mu$ lies between $\D_{-}+\mu$ and $E^f$.
 The system is insulating.
}
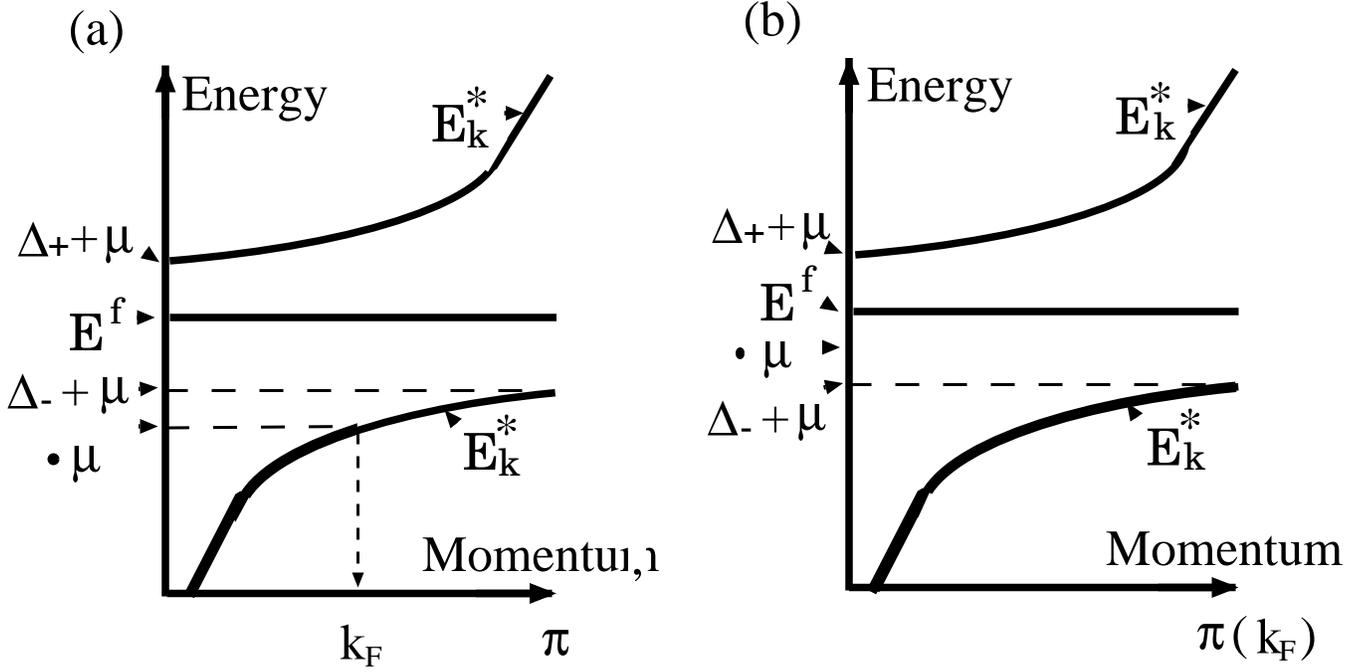
\item{ Fig. 2 \ : \ \ \ (a) \ Four point vertex
    $\G_{\p\p',\q}^{MM'.M''M'''}(\e,\e';\w)$.  (b) \ $k$-limit ( or $\w$-limit )
    four point vertex ${\hat \G}_{\p\p'}^k(\e,\e')$.
}
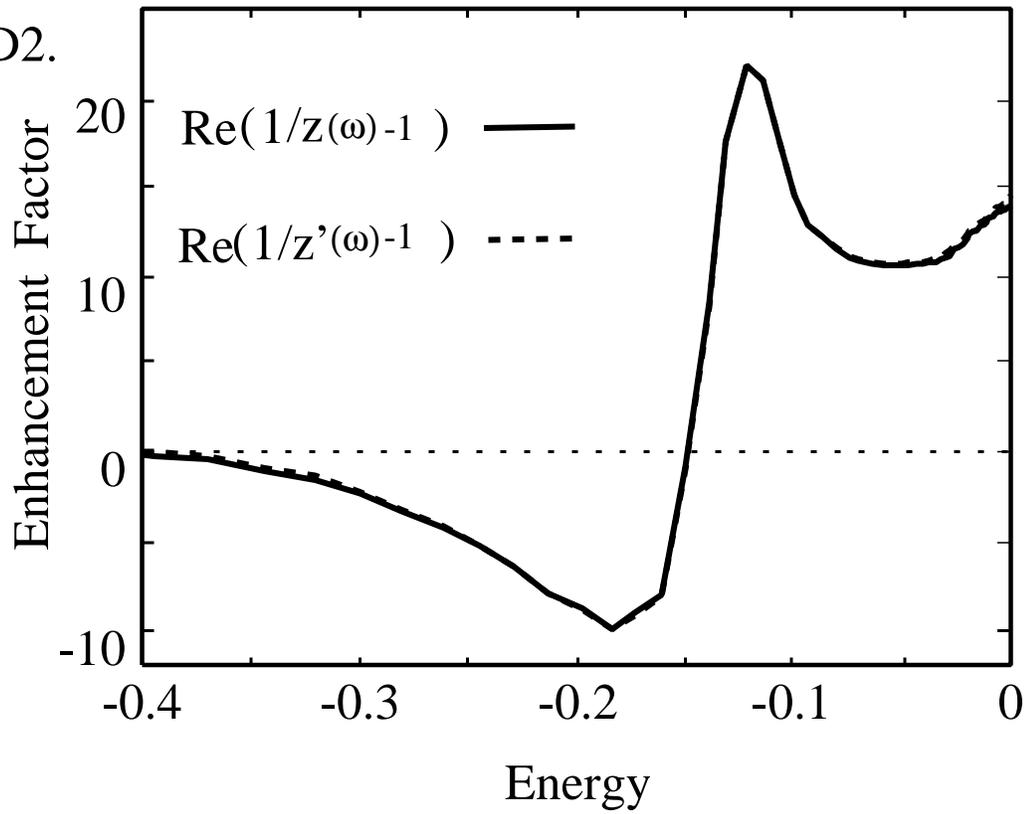
\item{ Fig. 3 \ : \ \ \ The schematic structure for $\chi_V$.
}
\item{ Fig. 4 \ : \ \ \ The structure of the vertex correction for the
    ( effective mass or magnetic ) enhancement factor in the
    $d=\infty$ approximation.  Here, $I$ represents the {\it local}
    irreducible four-point vertex, $\G^I$, and ${\hat \v}$ represents
    the particle-hole Green's functions, connecting between two sites.
    $l$, $m$, $n$ and $j$ represent the $f$-electron sites, on which
    we have to take summation.
}
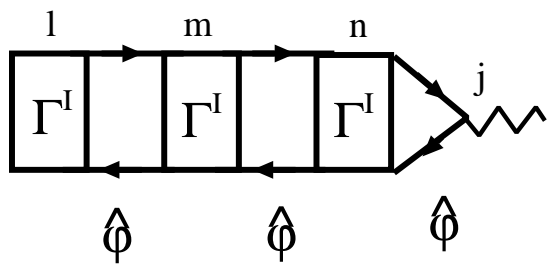
\item{ Fig. 5 \ : \ \ \ Two examples of $U^2$-order contributions for
    the susceptibility.  $j$, $l$, $m$ and $n$ represent the
    $f$-electron sites, on which we have to take summation.
}
\item{ Fig. 6 \ : \ \ \ $\G_{MM,M'M'}^I(\e,\e')$ represents the {\it
      irreducible} parallel four-point vertex with respect to the particle-hole
    pair.  In the $d=\infty$ limit, it is diagonal with respect to $M$
    and independent of $M$.
}
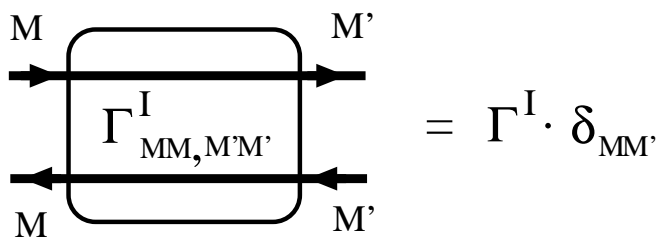
\item{ Fig. 7 \ : \ \ The definition for two kinds of the
    particle-hole Green's functions, $\v_a(\w)$ and $\v_b(\w)$.
    Both of them are independent of $M$.  
    $l$ represents the $f$-electron sites, on which we have
    to take summation.
    In (b), factor $M'$ comes from the Zeeman term.
}
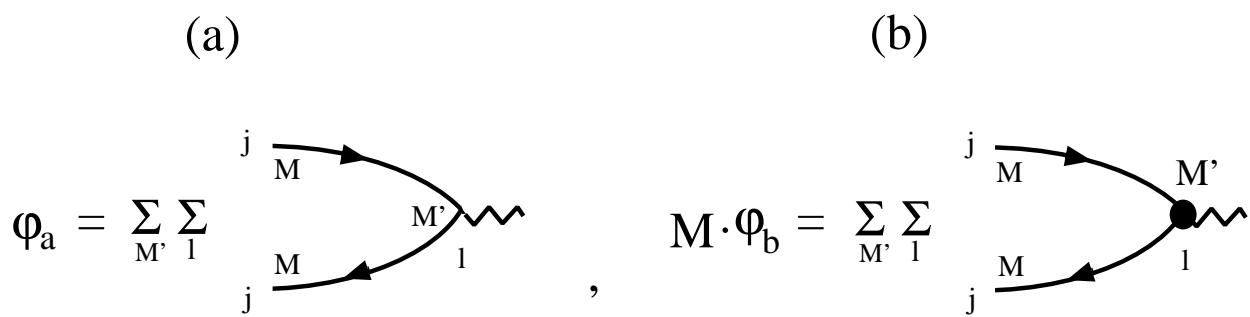
\item{ Fig. 8 \ : \ \ \ The schematic structure for the
    energy-derivative of the selfenergy.  This is composed of \{
    $\G^I$, $\v_a$ \}.
}
\item{ Fig. 9 \ : \ \ \ The first line represents the Bethe-Salpeter
    equation relating between $\G^{a'}$ and $\G^b$.  The second line
    is derived by using the first line twice.
}
\item{ Fig. 10 \ : \ \ \ The schematic structure for $\chi_V$.  This
    is composed of \{ $\G^I$, $\v_{a'}$ \}.  We stress that two
    enhancement factors appear in total on both sides of
    $Q_\gamma(\w)$ in the last line.  $\chi_V''$ turns out to vanish
    identically.
}
\item{ Fig. B$\cdot$1 \ : \ \ \ $\G(\e,\e')$ represents the {\it parallel}
    four-point vertex and $T(\e,\e')$ represents the {\it anti-parallel}
    four-point vertex, respectively.
}
\item{ Fig. C$\cdot$1 \ : \ \ \ Complex integration path $C$ in eq.
    (\ref{eqn:example2}).
}
\item{ Fig. D$\cdot$1 \ : \ \ \ The diagrams for the enhancement
    factors in SOPT.  The broken lines represent the Coulomb potential
    $U$.  (a) \ \{$1/z(\w)-1$\} in SOPT. This is constructed by the
    unperturbed $\v_a^0(\w)$.  (b) \ \{$1/z'(\w)-1$\} in SOPT. This is
    constructed by the unperturbed $\v_{a'}^0(\w)$.
}
\item{ Fig. D$\cdot$2 \ : \ \ \ Numerical results for the frequency
    dependence of the real parts of enhancement factors by SOPT, for
    the metallic case.  The line and the broken line represent
    \{$1/z(\w)-1$\} and \{$1/z'(\w)-1$\}, respectively.  The energy
    region for large enhancement in SOPT is about $\simle |\mu-E^f|$.
    We put $U^2= 1$, $3V^2= 0.16$, $E^f= -0.3$, $k_F= 0.885\pi$ and
    $\e_k= 2(|k|/\pi)^3-1$, respectively.
}
\item{ Fig. D$\cdot$3 \ : \ \ \ Numerical results for the frequency
    dependence of the real parts of enhancement factors by SOPT, for
    the insulating case.  The line and the broken line represent
    \{$1/z(\w)-1$\} and \{$1/z'(\w)-1$\}, respectively.  We put $U^2=
    1$, $3V^2= 0.25$, $E^f= -0.5$, $\mu= (\D_{-}+E^f)/2$ and $\e_k=
    2(|k|/\pi)^3-1$, respectively. 
}
%
%
\item{ Fig. D$\cdot$4 \ : \ \ \ The diagrams for the irreducible
    parallel four point vertex $\G^I(\e,\e')$ used in SC-SOPT.
}
\item{ Fig. D$\cdot$5 \ : \ \ \ Numerical results for the frequency
    dependence of the real parts of the enhancement factors by
    SC-SOPT, for the metallic case.  The line and the broken line
    represent \{$1/z(\w)-1$\} and \{$1/z'(\w)-1$\}, respectively.  The
    energy region for large enhancement in SC-SOPT is about $ \simle
    z(0)\cdot |\mu-E^f|$, which is renormalized compared with that by
    SOPT.  ( see Fig. D3. )  We put $U^2= 0.5$, $3V^2= 0.16$, $E^f=
    -0.3$, $k_F= 0.885\pi$ and $\e_k= 2(|k|/\pi)-1$, respectively.
}
\item{ Fig. E$\cdot$1 \ : \ \ The diagram for $1/z'(\w)$.  The first
    term is irreducible and the second term is reducible with respect
    to $U$.  As the second term includes $\chi_V''$, it vanishes
    identically.
}
\end{itemize}


\begin{thebibliography}{99}
\bibitem{Zou}
 Z. Zou and P. W. Anderson, Phys. Rev. Lett. {\bf 57} (1986) 2073.
\bibitem{Hanzawa_p}
 K. Hanzawa, Y. Yosida and K. Yamada, Prog. Theor. Phys. {\bf 81}
 (1989) 960.
\bibitem{paper1} H. Kontani and K. Yamada, J. Phys. Soc. Jpn. {\bf
    63} (1994) 2627.  \ \ \ In their paper, the theory of the AHE on
  the ferromagnetic metals proposed by ref. \cite{KL} is corrected and
  its reality is proved rigorously, by taking account of the
  correlation between electrons in a consistent way, which makes the
  lifetime of the quasiparticle finite.
%
\bibitem{KL} R. Karplus and J. M. Luttinger, Phys. Lev. {\bf 95}
  (1954) 1154.  \ \ \ The reality of the mechanism of the AHE proposed in
  their paper has been controversial for years :
see ref. \cite{Sm}.    
%
\bibitem{Sm}
 J. Smit, Physica {\bf 21} (1955) 877 ; {\bf 24} (1958) 39.
\bibitem{Zhang} F. C. Zhang and T. K. Lee, Phys. Rev. Lett. {\bf 58},
  (1987) 2728, G. Aeppli and C. M. Varma, Phys. Rev. Lett. {\bf 58},
  (1987) 2729,
 D. L. Cox, Phys. Rev. Lett. {\bf 58} (1987) 2730.
%
\bibitem{Hanzawa_v}
 K. Hanzawa, Y. Yosida and K. Yamada, Prog. Theor. Phys. {\bf 77}
 (1987) 1116.
%
\bibitem{insulator} For a brief review, see G. Aeppli and Z. Fisk,
  Comm. Cond. Mat. Phys. {\bf 16} (1992) 155.
%
\bibitem{Ohmi} For example, UPt$_3$ is frequently believed to shows
  the odd-parity superconducting state because of the constant Knight
  shift below the transition temperature ; see Y. Kohori, T. Kohara,
  H. Shibai, Y. Oda, Y. Kitaoka
 and K. Asayama, J. Phys. Soc. Jpn. {\bf 57} (1987) 395.
%
\bibitem{meanfield}
 S. M. M. Evans, J. Phys.: Condens. Matter {\bf 2} (1990) 9097. 
%
\bibitem{Anderson}
 P. W. Anderson and Z. Zou, Phys. Rev. Lett. {\bf 58} (1987) 2731.
%
\bibitem{Nakano} M. Nakano, Phys. Rev. B, {\bf 44} (1991) 10300.  \ \ 
  \ We note that $\chi_{\rm qp}$ and $\chi_{\rm nqp}$ defined in his
  paper are equal to
$\chi_P$ and $\chi_V$ introduced in our paper, respectively.
%
\bibitem{Luttinger}
 J. M. Luttinger and J. C. Ward, Phys. Rev. {\bf 118} (1960) 1417.
%
\bibitem{Shiba}
 H. Shiba, Prog. Theor. Phys. {\bf 54} (1975) 967.
%
\bibitem{Yamada_p}
 K. Yamada and K. Yosida, Prog. Theor. Phys. {\bf 76} (1986) 621.
%
\bibitem{RPA} By the RPA approximation, $\chi^{\rm RPA} =
  \chi^0/(1-U \rho(0)/6)$ and $\chi_V^{\rm RPA} = \chi_V^0/(1-U
  \rho(0)/(6\times2.19))$ are obtained straightforwardly.
 $\chi^0$ and $\chi_V^0$ is given by \S 4.
 When $U\simle 6/\rho(0)$, 
 $\chi^{\rm RPA} \simle \infty$ and $\chi_V^{\rm RPA}=1.84$.
%
\bibitem{d} W. Metzner and D. Vollhardt, Phys. Rev. Lett. {\bf 62}
  (1989) 324, U. Brandt and C. Mielsch, Z. Phys. B {\bf 75}
  (1989) 365, V. Janis, Z. Phys. B {\bf 83} (1991) 227, F. J. Ohkawa,
  Phys. Rev. B {\bf 44} (1991) 6812,
 A. Georgs and G. Kotliar, Phys. Rev. B {\bf 45} (1992) 6479.
%
\bibitem{paper3}
 H. Kontani and K. Yamada, in preparation.
%
\bibitem{Okada}
 K. Okada, K. Yamada and K. Yosida, Prog. Theor. Phys. {\bf 77} (1987) 
 1297.
%
\bibitem{AGD} A. A. Abrikosov, L. P. Gorkov and I. E. Dzyaloshinski,
  {\em Methods of Quantum Field Theory in Statistical Physics}
  ( Dover, New York, 1975 ).
%
\bibitem{Nozier} P. Nozi\`{e}res, {\em Theory of Interacting Fermi
    Systems}
  ( Benjamin, New York, 1964 ).
%
\bibitem{Kubo}
 R. Kubo, J. Phys. Soc. Jpn. {\bf 12} (1957) 570.
%
\bibitem{explain} The imaginary part of $-\ddw \Si_M(\w)$, given by
  (\ref{eqn:ddw_S}), is proportional to $\w$ for $|\w|\simle T^\ast$.
  Here, we denote the first term of the r.h.s of the first line in
  (\ref{eqn:ddw_S}) as $h_f(\w)$ and that of the second term as
  $h_c(\w)$, respectively.  
  We can show in the familiar FLT
  that the $\w$-linear coefficient of Im$\{ -\ddw \Si_M(\w) \}$ comes
  dominantly from that of Im$h_f(\w)$, which is about
  $3V^2/(\mu-E^f)^2$ $\sim$ $D^2/V^2$ times larger than that of $h_c(\w)$.  
  By use of the Kramers-Kronich relation, 
  the real part of $ -\ddw \Si_M(\w) $ is determined mainly
  by Im$\{ -\ddw \Si(\zeta) \}$ for
  $|\zeta|\simle T^\ast$.  Roughly speaking, $ -\ddw \Si(0)$ $\sim$
  $T^\ast \cdot \lim_{\w \rightarrow0}{\rm Im}( h_f(w)+ h_c(\w) )/\w$.
  Thus, we conclude that Re$\{ h_c(\w)/h_f(\w) \} \sim O(V^2/D^2) \ll
  1$ for
$|\w|\simle T^\ast$, and $h_c(\w)$ is negligible in this paper.
%
\bibitem{Storner} The polarization factor, $1-\G^{b,\b}(0,0) \cdot A
  z(0)\rho(0)$, is given up to the 1st order of $U$ as
$1+U\rho(0)/3$ ( $>1$ ).
%
\bibitem{explain2} For $J=5/2$ single-site Anderson model without ECF,
  whose Van Vleck susceptibility is rigorously zero, $R=(2J+1)/2J$ is
  proven rigorously in the same way as done in \S 6.
This is also expected for SU(6)-PAM without ECF under the approximation (\ref{eqn:symm}).
%
\bibitem{SUN} For instance, see A. J. Mills and P. A. Lee, Phys. Rev.
  B {\rm 35} (1987) 3394,
 B. Jin and Y. Kuroda, J. Phys. Soc. Jpn. {\bf 57} (1988) 1687.
%
\bibitem{Yosi} Such a property of the Pauli's principle has already
  been pointed out for orbital degenerate Anderson model : see
 A. Yoshimori, Prog. Theor. Phys. {\bf 55} (1976) 67.
%
\bibitem{Yamada_v}
 K. Yamada and M. Nakano, Prog. Theor. Phys. Suppl. {\bf 101} (1990) 419.
%
\bibitem{Schweitzer} H. Schweitzer and G. Czycholl, Z. Phys. B {\bf
    79} (1990) 377,
 H. Schweitzer and G. Czycholl, Z. Phys. B {\bf 67} (1991) 3724.
%
\bibitem{Muto} T. Mutou and D. S. Hirashima, J. Phys. Soc. Jpn.
  {\bf 63} (1994) 4475.
\end{thebibliography}
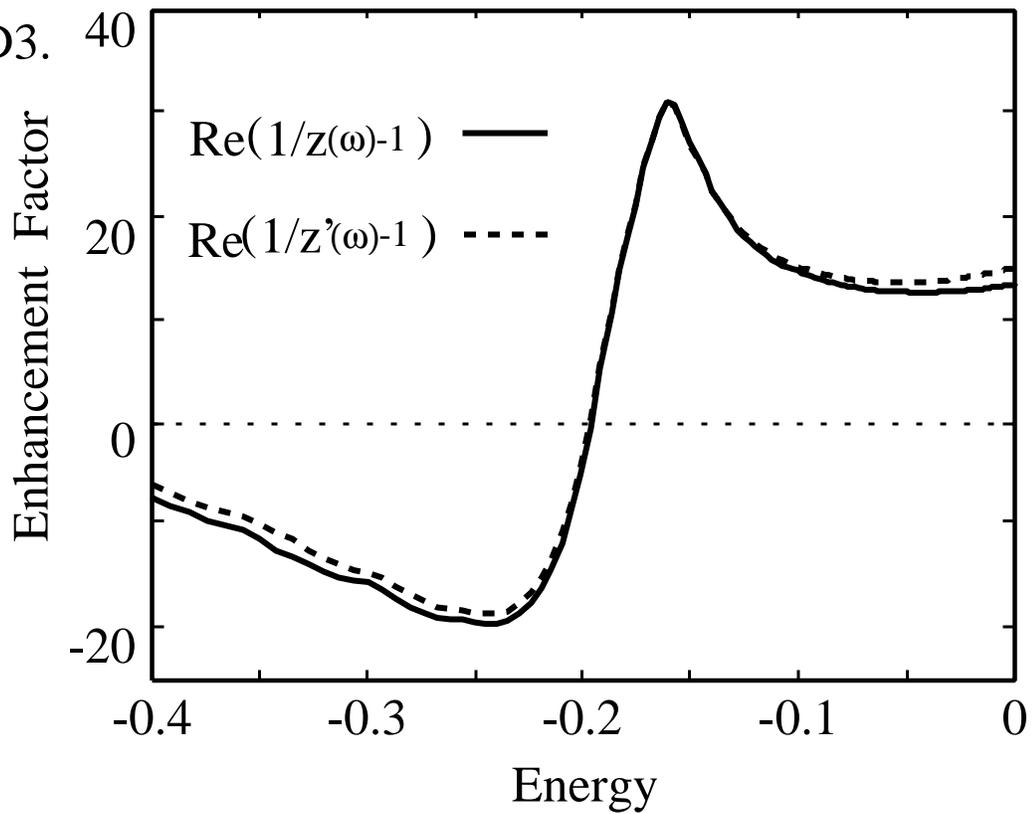
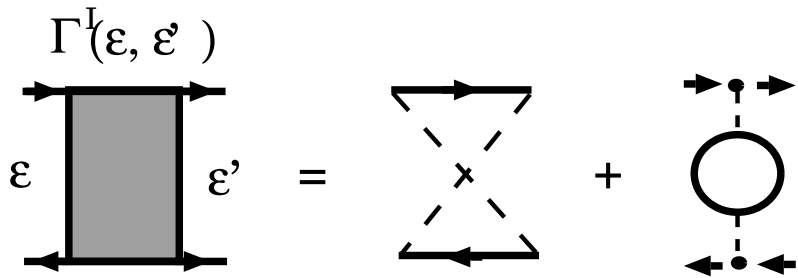
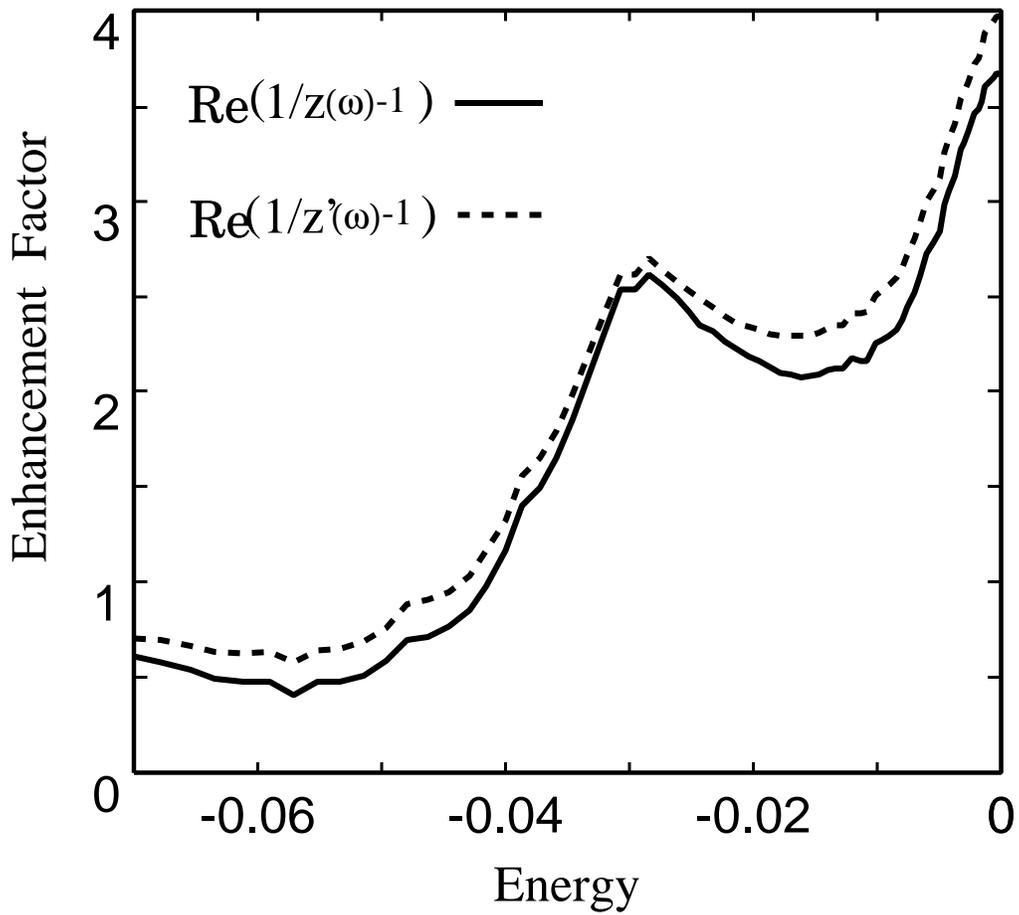
\end{document}